\documentclass[journal]{IEEEtran}

\usepackage{textcomp}
\usepackage{xcolor}
\usepackage{graphicx}
\usepackage{amsmath,amssymb,amsfonts}
\usepackage{algorithm}
\usepackage{algorithmic}
\usepackage{cite}
\usepackage{lipsum}
\usepackage{epstopdf}
\usepackage{stfloats}
\usepackage{bm}
\usepackage{xcolor}
\usepackage{colortbl,booktabs}

\begin{document}
\title{Towards Reliable UAV-Enabled Positioning in Mountainous Environments: System Design and Preliminary Results}

\author{Zijie~Wang,
        Rongke~Liu,~\IEEEmembership{Senior Member,~IEEE,}
        Qirui~Liu,
        Lincong~Han
\thanks{This work was supported by the Beijing Municipal Science and Technology Project (Z181100003218008).}
\thanks{Z. Wang and R. Liu are with the School
of Electronic and Information Engineering, Beihang University, Beijing 100191,
China (e-mail: wangmajie@buaa.edu.cn; rongke\_liu@buaa.edu.cn).}
\thanks{Q. Liu is with the School
of Electronic and Information Engineering, Beihang University, Beijing 100191,
China.}
\thanks{L. Han is with the School
of Electronic and Information Engineering, Beihang University, Beijing 100191,
China.}
}

\markboth{Journal of \LaTeX\ Class Files,~Vol.~14, No.~8, August~2015}%
{Shell \MakeLowercase{\textit{et al.}}: Bare Demo of IEEEtran.cls for IEEE Journals}

\maketitle

\begin{abstract}
Reliable positioning services are extremely important for users and devices in mountainous environments as it enables a variety of location-based applications. However, in such environments, the service reliability of conventional wireless positioning technologies is often disappointing. Frequent non-line-of-sight (NLoS) propagation and poor geometry of available anchor nodes are two significant challenges. Due to the high maneuverability and flexible deployment of unmanned aerial vehicles (UAVs), UAV-enabled positioning could be a promising solution to these challenges. Compared with satellites and terrestrial base stations, UAVs are capable of flying to places where both the propagation conditions and geometry are favorable for positioning. The eventual aim of this research project is to design a novel UAV-enabled positioning system that uses a low-altitude UAV platform to provide highly reliable services for ground users in mountainous environments. In this article, we introduce the recent progress made in the first phase of our project, including the following. First, the structure of the proposed system and the positioning method used are determined after comprehensive consideration of various factors. Utilizing the digital elevation model of the realistic terrain, we then establish a geometry-based NLoS probability model so that the NLoS propagation can be treated as a type of fault during the reliability analysis. Most importantly, a reliability prediction method and the corresponding metric are developed to evaluate the system's ability to provide reliable positioning services. At the end of this article, we also propose a voting-based method for improving the service reliability. Numerical results demonstrate the tremendous potential of the proposed system in reliable positioning.
\end{abstract}


\IEEEpeerreviewmaketitle

\section{Introduction}
\subsection{Motivation}
\IEEEPARstart{M}{otivated} by the explosive growth of applications that require or benefit from location information, positioning technologies are playing an increasingly important role in our everyday lives \cite{LBS_Intr_1}. Both the fifth generation (5G) wireless network and the Narrowband Internet of Things (NB-IoT) have considered positioning as an enabling technology and essential service in 3rd Generation Partnership Project (3GPP) Release 16 and Release 14, respectively \cite{TR_38.855,IoT_LBS}. Mountainous environment is one of the typical deployment scenarios of 5G and NB-IoT \cite{5G_Mount,IoT_Mount}, in which the positioning service is also extremely important. Different from plain or city environments, the terrain in mountainous environments is complex and highly variable \cite{Mount_Terr}. There may be several completely different landforms like peaks, steep cliffs and deep valleys within a small area, so that large position errors caused by faults or anomalies may result in heavy economic losses or even loss of life. Thus, the reliability of positioning services should be a major concern for location-based applications in mountainous environments, especially those life-critical applications like field rescue and disaster management \cite{Pos_Disa}.
\begin{figure}[!t]
\centering
\includegraphics[height=3.08in,width=3.08in]{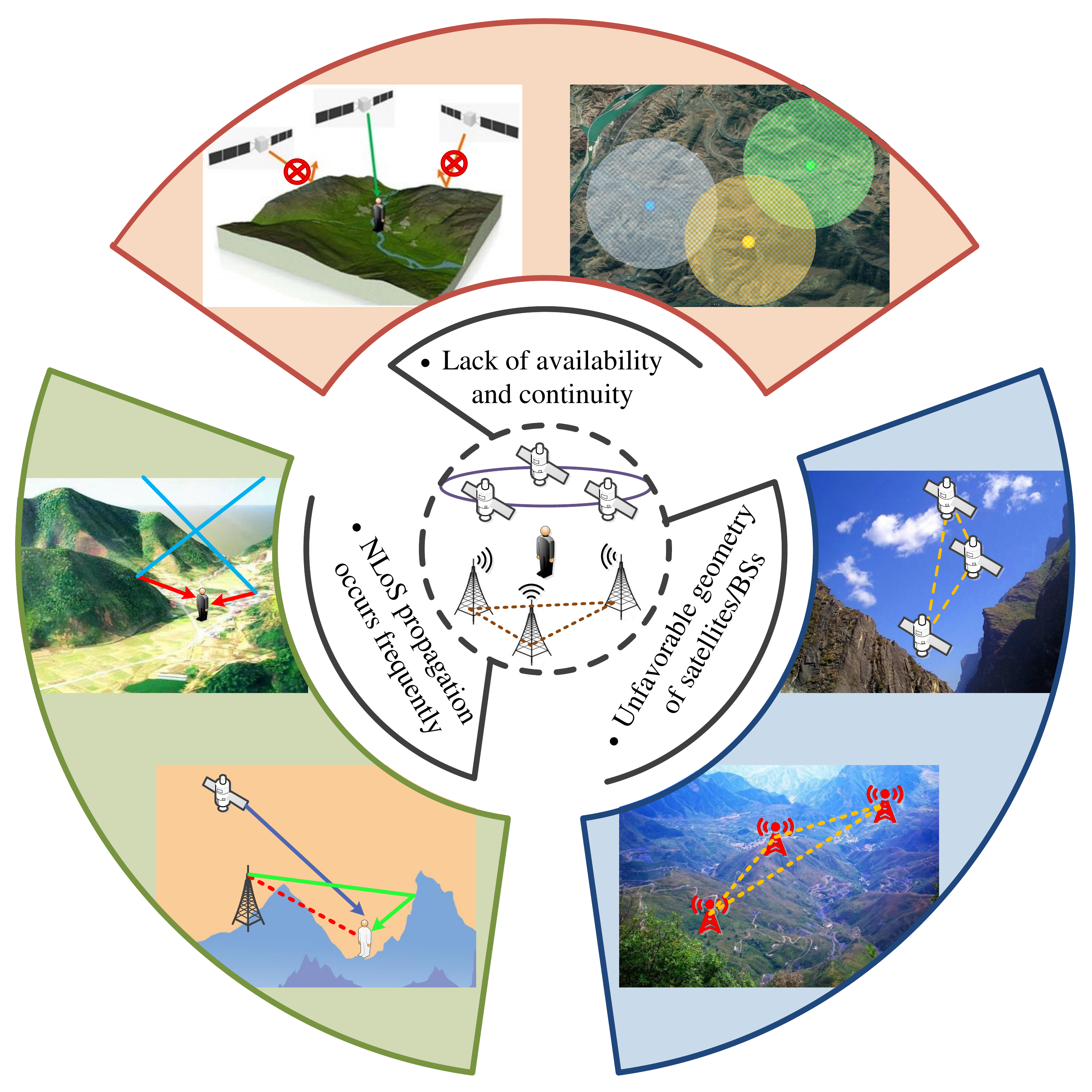}
\caption{Limitations of existing positioning technologies in mountainous environments.}
\label{fig_1}
\end{figure}

Thanks to the continuous development of consumer electronics industries over the past two decades, almost every smartphone or wearable device today is equipped with a global navigation satellite system (GNSS) receiver, which could provide users with satisfactory positioning service in some open-sky environments \cite{Phone_GNSS_1,Phone_GNSS_2}. Besides, since 3GPP Release 9, terrestrial cellular networks have supported a variety of cellular-based positioning technologies like the observed time difference of arrival (OTDoA) positioning \cite{OTDoA_Rel9}. On account of this, mobile phone users could use the cellular network to locate themselves even in some GNSS degraded environments like dense urban and indoors \cite{LTE_Urban,LTE_Indoor}. However, there is a question that should be asked and seriously considered: Are existing wireless positioning technologies, including GNSS systems and the cellular-based positioning, sufficient to ensure the reliability of positioning services in mountainous environments? Unfortunately, the answer to this question is no. As shown in Fig. 1, for the widely used GNSS positioning technology, the weak signals are frequently blocked by rugged terrain like peaks and ridges, resulting in very limited availability and continuity of service \cite{GNSS_Block,GNSS_Num}. Moreover, the signal detected by the receiver may not be the desired direct (LoS) signal from the satellite, but the signal reflected from mountains \cite{GNSS_NLoS}. This phenomenon, known as NLoS propagation, will severely degrade the performance of positioning service, causing errors of several meters to hundreds of meters. Another challenge for GNSS users in mountainous environments is the unfavorable geometry of the satellites in-view \cite{GNSS_Geo}. For instance, the satellites available to the user in a valley are mainly in the along-valley direction, since the signals with line of sight going across the valley are very likely to be blocked. This poor satellite geometry leads to unsatisfactory accuracy and reliability in the across-valley direction. Similar to GNSS positioning, cellular-based positioning has the same limitations \cite{LTE_Chal}, including insufficient number of available base stations (BSs), unfavorable geometry of BSs, and the NLoS propagation phenomenon, making it unsuitable for reliable positioning. To make matters worse, since the elevation angles of terrestrial BSs are generally much lower than those of GNSS satellites, cellular signals experience more NLoS propagation than GNSS signals \cite{LTE_NLoS}. It is therefore unwise to expect cellular networks to perform better than GNSS systems. Up to now, the implementation of reliable positioning in mountainous environments remains an open question.
\begin{figure}[!t]
\centering
\includegraphics[height=2.00in,width=3.05in]{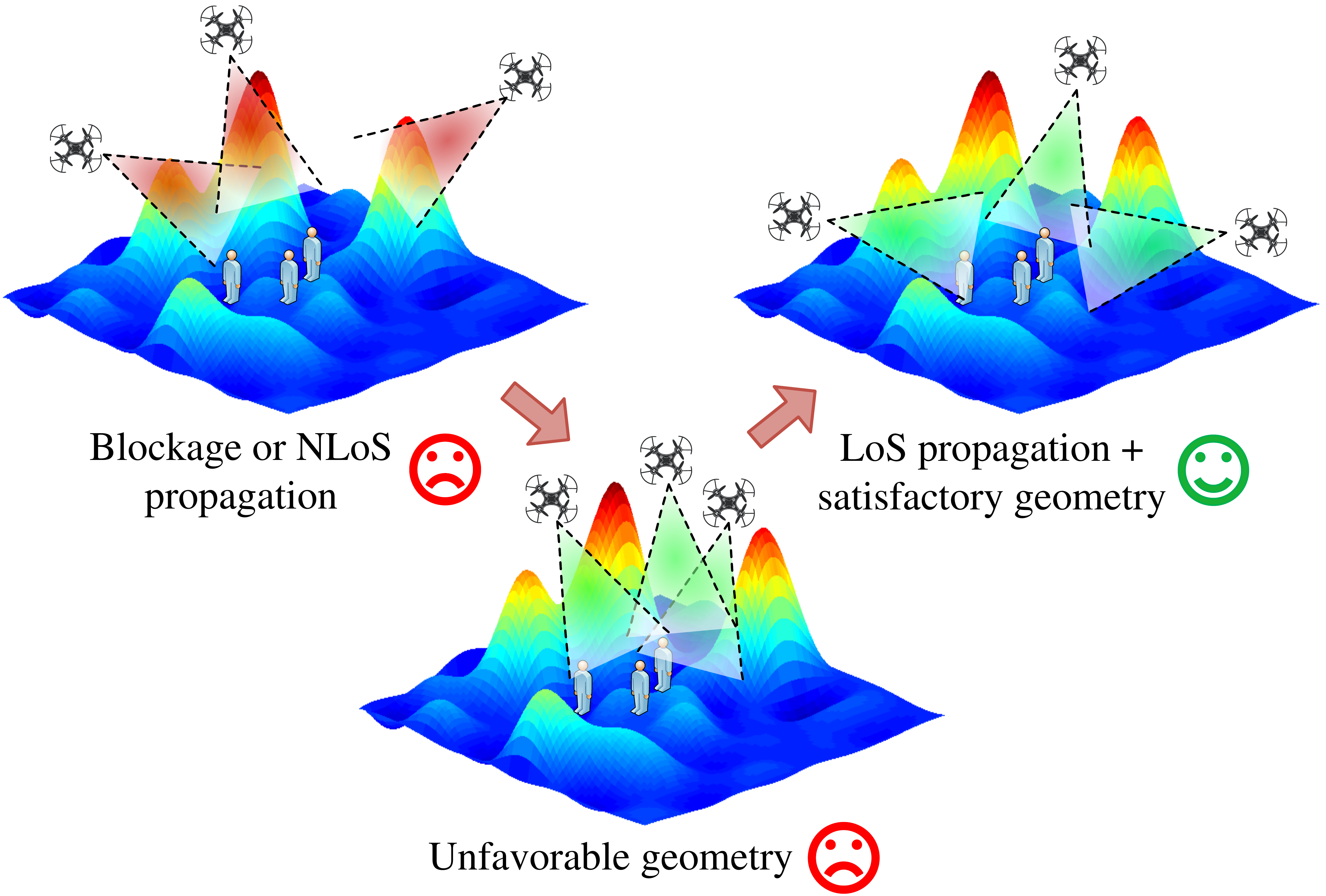}
\caption{Advantages of UAV-enabled positioning in mountainous environments.}
\label{fig_2}
\end{figure}

With their high maneuverability and flexible deployment, the UAV platform could be a promising solution to the aforementioned challenges in mountainous environments. Utilizing high-performance airborne navigation equipment like real-time kinematic (RTK) receivers, UAVs have the ability to precisely determine their own locations in real time, enabling them to be employed as anchor nodes for positioning \cite{UAV_RTK}. As shown in Fig. 2, compared with satellites operating in fixed orbits or BSs fixed on the ground, UAV platforms have their unique advantages. First, UAVs are capable of flying to places where the channel conditions to users are good \cite{UAV_LoS}. Therefore, through the optimization of the flight trajectory or deployment strategy, UAV platforms can provide users with a sufficient number of measurements, and effectively reduce the probability of NLoS propagation. Moreover, if the geometry of anchor nodes is also taken into account in the design of the UAV trajectory \cite{UAV_IoT}, the problem of unfavorable geometry common in GNSS and cellular-based positioning could be largely avoided. From the above analysis, it can be concluded that the UAV-enabled positioning system is a good choice for providing reliable positioning services in mountainous environments. Thus, the eventual aim of this research project is to design such a system.

\subsection{Related Work}
UAV-enabled positioning that employs UAV platforms to provide positioning services is not a new concept. In fact, some studies pointed out more than a decade ago that UAVs and many other types of platforms have the potential to be used as mobile anchor nodes to determine users' locations \cite{Old_UAV_1,Old_UAV_2}. In the past few years, several prototypes of UAV-enabled positioning system have been developed and tested in the research community, such as HAWK (UMass Lowell) \cite{UAV_HAWK} and GuideLoc (NWU, China) \cite{UAV_GuideLoc}. Most of these existing systems adopt range-free localization approaches, which limits their ability to provide accurate and reliable positioning services. Moreover, their positioning performance is tested in some ideal environments like sports field or campus, which cannot effectively reflect the accuracy and reliability of positioning services in practical applications. With the development of airborne sensors, many recent studies began to use range-based localization approaches, so as to make their systems suitable for missions requiring high positioning accuracy. Sallouha \emph{et al}. \cite{UAV_RSS_1,UAV_RSS_2} proposed a system that uses UAVs and received signal strength (RSS) technique to locate terrestrial nodes, and analyzed the influences of UAV altitude and trajectory on positioning performance. In \cite{UAV_IoT}, a UAV platform was employed to assist the terrestrial IoT network in energy-efficient data collection and three-dimensional (3-D) TDoA positioning. These two studies mainly focused on the improvement of positioning accuracy, and used the Cram\'er-Rao lower bound (CRLB) to evaluate the system performance. The CRLB is a useful evaluation metric for location estimation under fault-free conditions as it defines a lower limit for the variance of any unbiased estimator. However, the CRLB cannot reflect the positioning accuracy and reliability under faulty conditions where most location estimators are no longer unbiased. So far, it seems that the service reliability in complex environments has not become a research hotspot in the field of UAV-enabled positioning.

Unlike UAV-enabled positioning, the GNSS community has paid a great deal of attention to reliability issues. The detection of service failures caused by faults or anomalies is one of the most basic capabilities for reliable positioning systems. In GNSSs, Receiver Autonomous Integrity Monitoring (RAIM) is a well-known technique used for detecting large position errors \cite{RAIM_Intr_1}. Since the late 1980s, many different RAIM methods have been developed and successfully applied in the aviation domain, such as the least-squares-residuals (LS) method and the solution separation (SS) method \cite{RAIM_Intr_1,RAIM_SS,RAIM_PS}. It should be noted that these classic RAIM methods are based on the single-fault assumption, which is not true for UAV-enabled positioning in mountainous environments. First, there is no doubt that the failure rates of low-cost airborne sensors are much higher than those of GNSS satellites. Moreover, frequent NLoS propagation also increases the probability of multiple simultaneous faults. In \cite{Multi_Fault_1,Multi_Fault_2}, the LS and SS RAIM were extended to multiple-fault conditions. However, as one of the major causes of service failures in mountainous environments, the NLoS propagation was not considered in these studies \cite{RAIM_NLoS}. Thus, the RAIM technique used in GNSSs cannot be transported directly into UAV-enabled positioning, but it still provides valuable inspiration for the design of our reliable positioning system.

\subsection{Problems and Contributions}
As mentioned above, our aim is to design a UAV-enabled positioning system to provide reliable services in mountainous environments. Specifically, we plan to achieve reliable positioning by solving the following problems.
\begin{enumerate}
\item \emph{Reliability prediction}: prediction of the minimum position error that can be effectively detected during the positioning process. If the predicted minimum detectable error is less than or equal to the user’s maximum tolerable error, then the UAV is allowed to take off and begin to provide positioning services.
\item \emph{Reliability enhancement}: adjustment of the UAV trajectory or deployment strategy when the above condition is not met. The objective of reliability enhancement is to reduce the minimum detectable error below the tolerable limit so that each service failure can be detected.
\item \emph{Reliability evaluation and maintenance}: detection and handling mechanisms for service failures occurring during the positioning process. In the presence of service failures, the system should be able to detect them and mitigate their impacts on positioning service through approaches like fault detection and exclusion (FDE), so as to maintain the error within an acceptable range.
\end{enumerate}

It can be foreseen that the study of the above three problems would be a large research project requiring a lot of time and effort. Thus, we decide to divide this project into multiple phases and carry out them in turn. In the first phase which has just been completed, the reliability prediction problem have been fully studied and solved. In addition, we have also conducted a preliminary study on reliability enhancement.

This article presents the progress we have made in the first phase of our project. Specifically, the main contributions of this article are summarized as follows.
\begin{itemize}
\item \emph{System design}: the structure and operation scheme of the proposed UAV-enabled positioning system is determined. The user’s location is estimated from range measurements obtained by a low-altitude UAV platform at multiple ``service points''. We choose the two-way ranging based positioning as the positioning method used in the proposed system and establish the corresponding measurement error model.
\item \emph{Modeling of service failures}: in this article, we consider two major causes of service failures in the proposed system. One of them is called ``internal faults'', which includes various anomalies in airborne sensors or user equipment, and the other is the well-known NLoS propagation phenomenon. The impacts of these two types of faults on positioning performance are theoretically analyzed and modeled. In particular, we develop a geometry-based model for calculating the prior probability of NLoS propagation, which is more conducive to reliable positioning than the conventional distance-based or elevation angle based models.
\item \emph{Proposal of a reliability prediction method}: based on the failure model mentioned above, we propose a reliability prediction method and derive the corresponding metric. This method takes account of the multiple-fault conditions common in mountainous environments, making it suitable for our system. The minimum detectable error during the positioning process is derived and regarded as the metric for evaluating reliability. The proposed method enables the system operator to predict the service reliability before the UAV takes off, thereby facilitating decision-making.
\item \emph{Preliminary study on reliability enhancement}: it is found in simulation that for a given set of service points, there may be some ``hazardous areas'' in which the predicted reliability does not meet the mission's requirement. Thus, we further study the problem of reliability enhancement. We first use the 8-connected neighborhood to segment the prediction result and extract hazardous areas. Then, we propose a voting-based method to analyze the cause of each hazardous area and provide guidance for adjusting the location of service points. Numerical results show that the proposed method is very helpful for improving the service reliability.
\end{itemize}

To the best of our knowledge, this work is the first to study the service reliability of UAV-enabled positioning systems in mountainous environments.

\subsection{Organization and Notations}
This article is organized as follows. The structure of the proposed system and some basic models are given in Section II. In Section III, the proposed reliability prediction method and its corresponding metric are derived. Section IV introduces our preliminary study on reliability enhancement. Section V provides numerical results to demonstrate the tremendous potential of our system and the effectiveness of the proposed methods. Finally, Section VI concludes this article.

The main notations used in this article are summarized as follows. Scalars are denoted by italic letters ($a$). Column vectors and matrices are denoted by lowercase and uppercase boldface letters (${\bf{a}}$ and ${\bf{A}}$), respectively. The superscript $T$ indicates the transpose operation (${{\bf{A}}^T}$) and superscript $ - 1$ indicates matrix inverse (${{\bf{A}}^{ - 1}}$). $\left\|  \cdot  \right\|$ represents the Euclidean norm of a vector. ${\bf{I}}$ is the identity matrix. $\left\langle  \cdot  \right\rangle $ denotes the rounding operation.

\section{System Design}
\begin{figure}[!t]
\centering
\includegraphics[height=1.75in,width=3.45in]{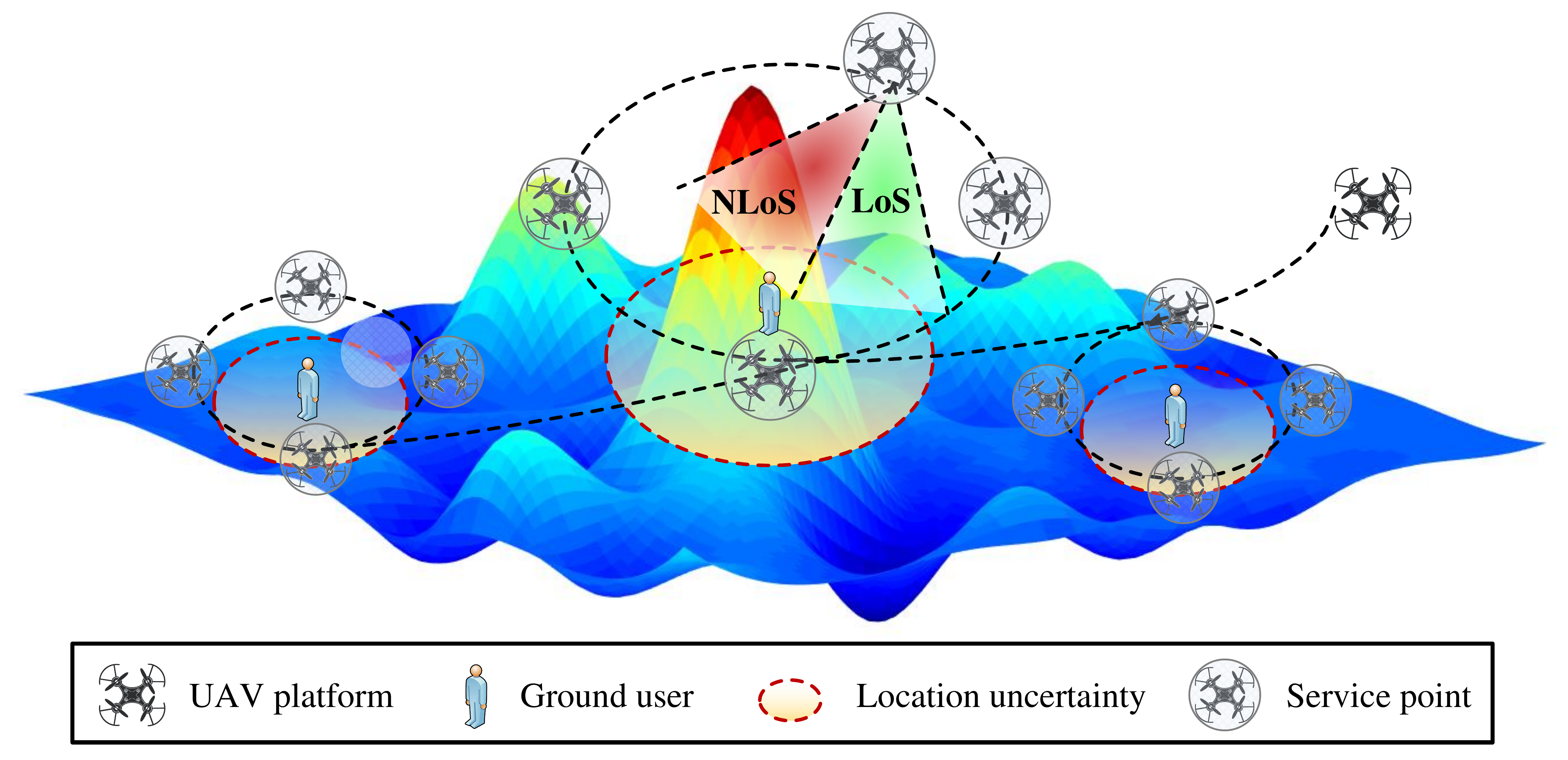}
\caption{Proposed UAV-enabled positioning system.}
\label{fig_3}
\end{figure}

\begin{figure}[!t]
\centering
\includegraphics[height=1.40in,width=2.73in]{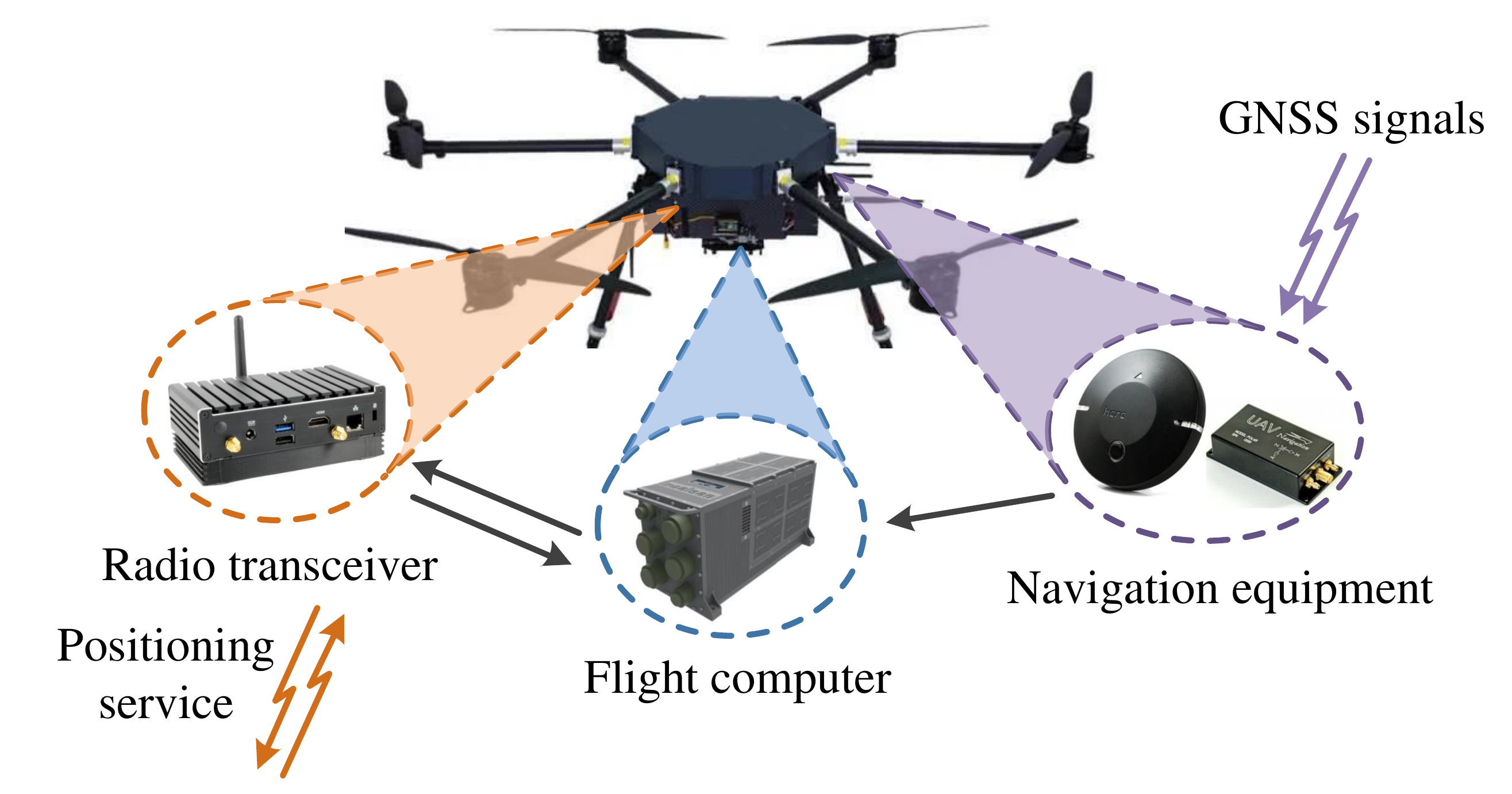}
\caption{Main components of the UAV platform.}
\label{fig_4}
\end{figure}

In this project, as shown in Fig. 3, we consider a scenario consisting of a low-altitude UAV platform and multiple ground users who scattered in a typical mountainous environment. The users could be hikers or injured people who are trapped and awaiting rescue, and our first priority is to determine their locations. Each user has a barometer embedded in his/her smartphone that could provide accurate altitude information. However, in such an environment, users cannot use the GNSS systems or cellular networks to obtain their horizontal locations due to the lack of service availability and reliability. Therefore, we expect the UAV platform to undertake the task of locating ground users. Fig. 4 shows the three main types of payloads carried on the UAV platform: 1) the navigation equipment used to determine the real-time precise location of the UAV itself; 2) a radio transceiver for transmitting and receiving positioning reference signal (PRS); and 3) a flight computer for flight control and location estimation. These payloads enable the UAV platform to be employed as an aerial anchor node to provide positioning services for ground users.

The UAV platform flies at a fixed altitude $h_B$. During the mission, the UAV flies from the initial location to the final location along a specific trajectory under control of the flight computer. When passing through the area where a ground user is located, the UAV will hover at multiple carefully selected “service points (SPs)” and perform two-way ranging (TWR). After all the SPs corresponding to the user have been traversed, its location can be estimated from the obtained range measurements utilizing multilateration algorithms.

In this article, we mainly focus on the reliability prediction and enhancement of the positioning service provided for each user. Besides, we assume that ground users are far apart, so that the UAV platform can only serve one user at each SP. Thus, the overall scenario described above can be divided into several single-user scenarios, as shown in Fig. 5. The precise location of the user is unknown before the UAV provides the positioning service. However, the location uncertainty area where the user is located can be obtained in advance based on historical information and motion prediction. The boundary of the uncertainty area is indicated by the red dotted line in Fig. 5, and its projection onto the horizontal plane is a circle of radius ${R_{Un}}$. For convenience, the location uncertainty area is discretized into $M$ sample points, denoted by the set ${\cal M} \buildrel \Delta \over = \left\{ {1,2, \cdots ,M} \right\}$. The 3-D location of the m-th sample point is denoted by the horizontal coordinate ${{\bf{u}}_m} = {\left( {x_U^m,y_U^m} \right)^T} \in {\mathbb{R}^{2 \times 1}}$, $m \in {\cal M}$, and the altitude $h_U^m$. In addition, the prior probability that the user is located at each sample point is assumed to be equal. For each ground user, we set $K$ SPs to provide positioning service, denoted by the set ${\cal K} \buildrel \Delta \over = \left\{ {1,2, \cdots ,K} \right\}$. The horizontal coordinate of the k-th SP is denoted by ${{\bf{w}}_k} \buildrel \Delta \over = {\left( {x_B^k,y_B^k} \right)^T} \in {\mathbb{R}^{2 \times 1}}$, $k \in {\cal K}$. It should be noted that there is a minimum distance requirement for the setting of SPs. Specifically, the horizontal distance between a SP and the center of the location uncertainty area should be no less than ${d_{\min }}$. The reason for this requirement is to ensure that large position errors caused by faults or anomalies will not bring significant changes to the geometry of SPs relative to the user, so that the service reliability could be evaluated or predicted without making assumptions on the values or distributions of unknown faults. Obviously, reducing the horizontal distance will increase the UAV elevation angle, resulting in a smaller probability of signal blockage or NLoS propagation. Thus, in this article, in order to ensure the reliability of the positioning service, SPs will be set on a circle of radius ${d_{\min }}$, which is indicated by the black dotted line in Fig. 5. Since the optimal location of SPs may be outside this circle if other factors like energy consumption are considered in future work, we retain the name ${d_{\min }}$, and hope that it will not cause inconvenience to the readers' understanding.
\begin{figure}[!t]
\centering
\includegraphics[height=1.43in,width=2.80in]{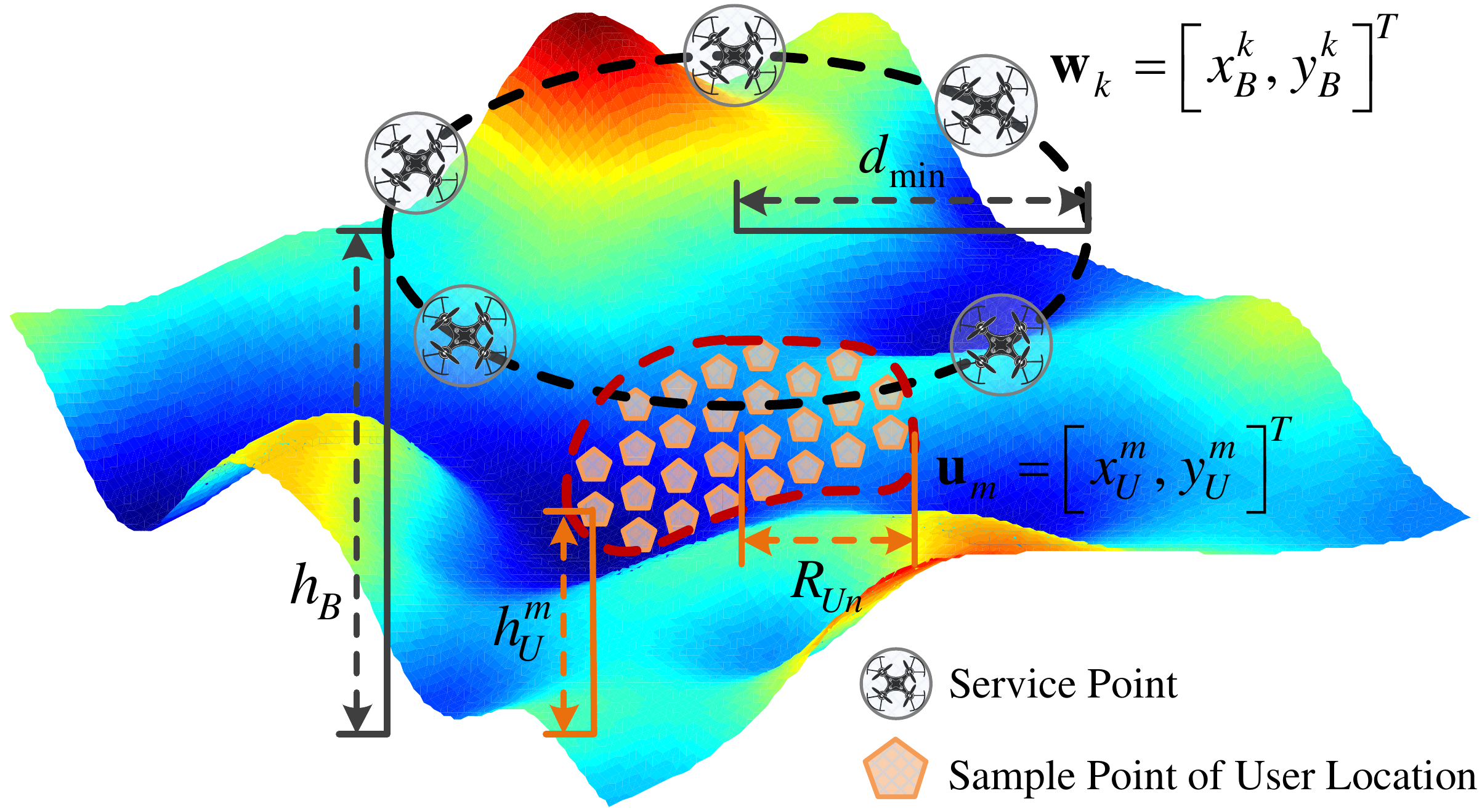}
\caption{Scenario of single user.}
\label{fig_5}
\end{figure}

The above paragraphs describe the structure, operation scheme and application scenario of the proposed UAV-enabled positioning system. In the following subsections, we provide some technical details. We first introduce a geometry-based probability model used to characterize the propagation conditions in mountainous environments. Then, the TWR-based positioning method and the corresponding measurement error model are analyzed in detail. Finally, the probability model of service failure that is essential for reliability prediction is established.

\subsection{Signal Propagation Model}
\begin{figure}[!t]
\centering
\includegraphics[height=1.23in,width=3.45in]{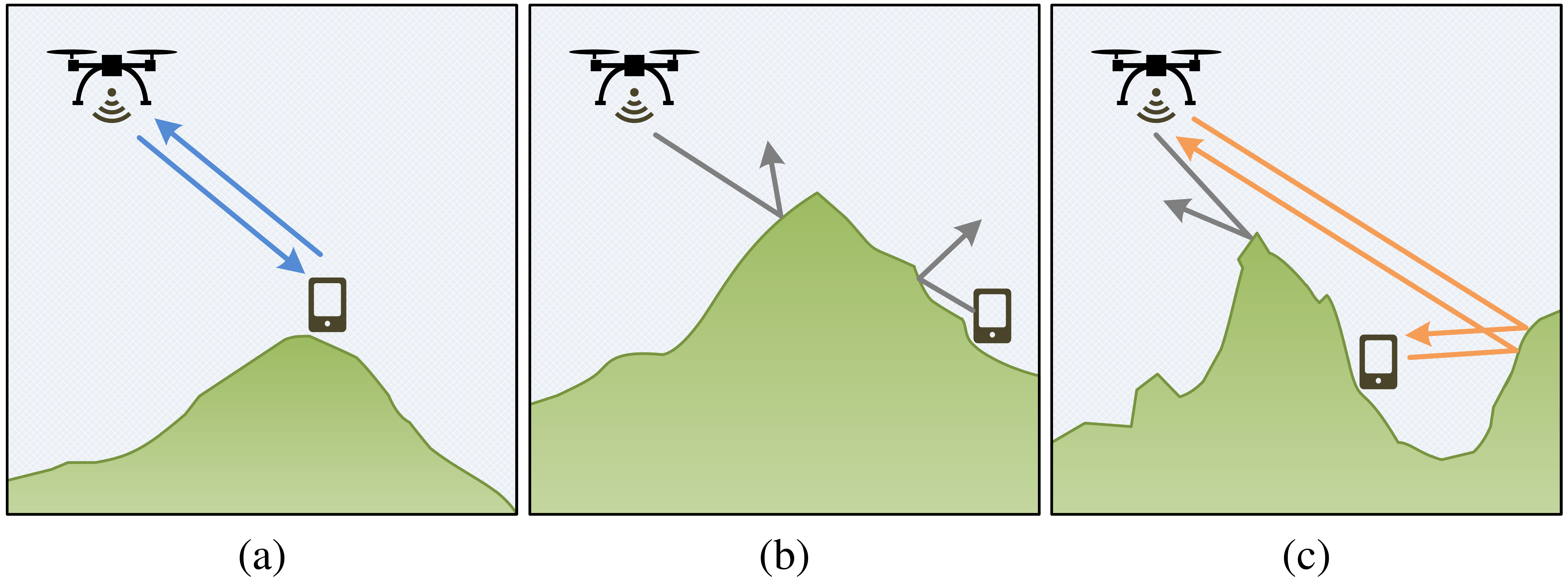}
\caption{Three types of propagation conditions in mountainous environments: (a) LoS condition, (b) Blockage condition and (c) NLoS condition.}
\label{fig_6}
\end{figure}

As shown in Fig. 6, for air-to-ground (A2G) channels in mountainous environments, there are three types of propagation conditions that need to be analyzed: 1) \emph{LoS condition} where a LoS radio path exists between the UAV and the ground user; 2) \emph{Blockage condition} where both the direct (LoS) and reflected signals are blocked by surrounding terrain and cannot be detected by the receiver; and 3) \emph{NLoS condition} where the direct signal is blocked and only reflected signals are received. Among these three conditions, the LoS condition is the most favorable one since it provides range measurement with acceptable error. The occurrence of the blockage condition will reduce the number of available measurements and thus affect the positioning performance, but it is not a direct cause of service failures. The NLoS condition is always a challenging problem for most range-based positioning systems as it imposes significant bias in both ranging and positioning. It should be noted that the multipath interference that frequently occurs in mountainous environments is not considered in this article, because we believe that existing techniques such as the multipath estimating delay lock loop (MEDLL) are sufficient to mitigate the impacts of multipath \cite{LTE_MEDLL}. From the above analysis, it can be concluded that these three propagation conditions have completely different effects on positioning services, requiring us to treat them differently. However, the exact propagation condition of an A2G channel cannot be fully determined before the wireless link is established. Thus, what we really want to achieve in this subsection is the determination of the prior probability of each condition.

Different from most existing research that employs distance-based or elevation angle based LoS probability models to characterize the A2G channel \cite{TR_38.901,Angle_based}, in this article, we develop a geometry-based stochastic model based on the DEM of the realistic terrain. Fig. 7 shows the terrain profile between the k-th SP of the UAV platform and the m-th sample point of user's location, which is reconstructed using the DEM data. The methods for reconstructing the terrain profile are described in \cite{DEM_Rec}. Assuming that the measurement error in DEM data is negligible and DEM resolution is extremely high, the LoS condition could be easily identified by checking whether all DEM data points are below the UAV-user LoS (blue solid line). Obviously, this assumption is unrealistic in practice. To address this problem, we model the ``terrain uncertainty'' caused by measurement error and limited resolution as an additive white Gaussian noise ${n_h}$ (orange dotted line) added to the minimum difference in height $\Delta h_{k,\min }^m$ between DEM data points and the UAV-user LoS.\footnote{Before calculating $\Delta h_{k,\min }^m$, those DEM data points within 20 meters of the ground user are excluded in advance. The reason for doing this is to avoid the unrealistic situation where the propagation condition is only determined by the DEM data point closest to the user.} Then, the actual minimum height difference with the consideration of location uncertainty is given by
\begin{equation}
\Delta \tilde h_{k,\min }^m = \Delta h_{k,\min }^m + {n_h},\quad{\rm{  }}{n_h} \sim {\cal N}\left( {0,\sigma _h^2} \right),
\end{equation}
\begin{figure}[!t]
\centering
\includegraphics[height=1.50in,width=2.69in]{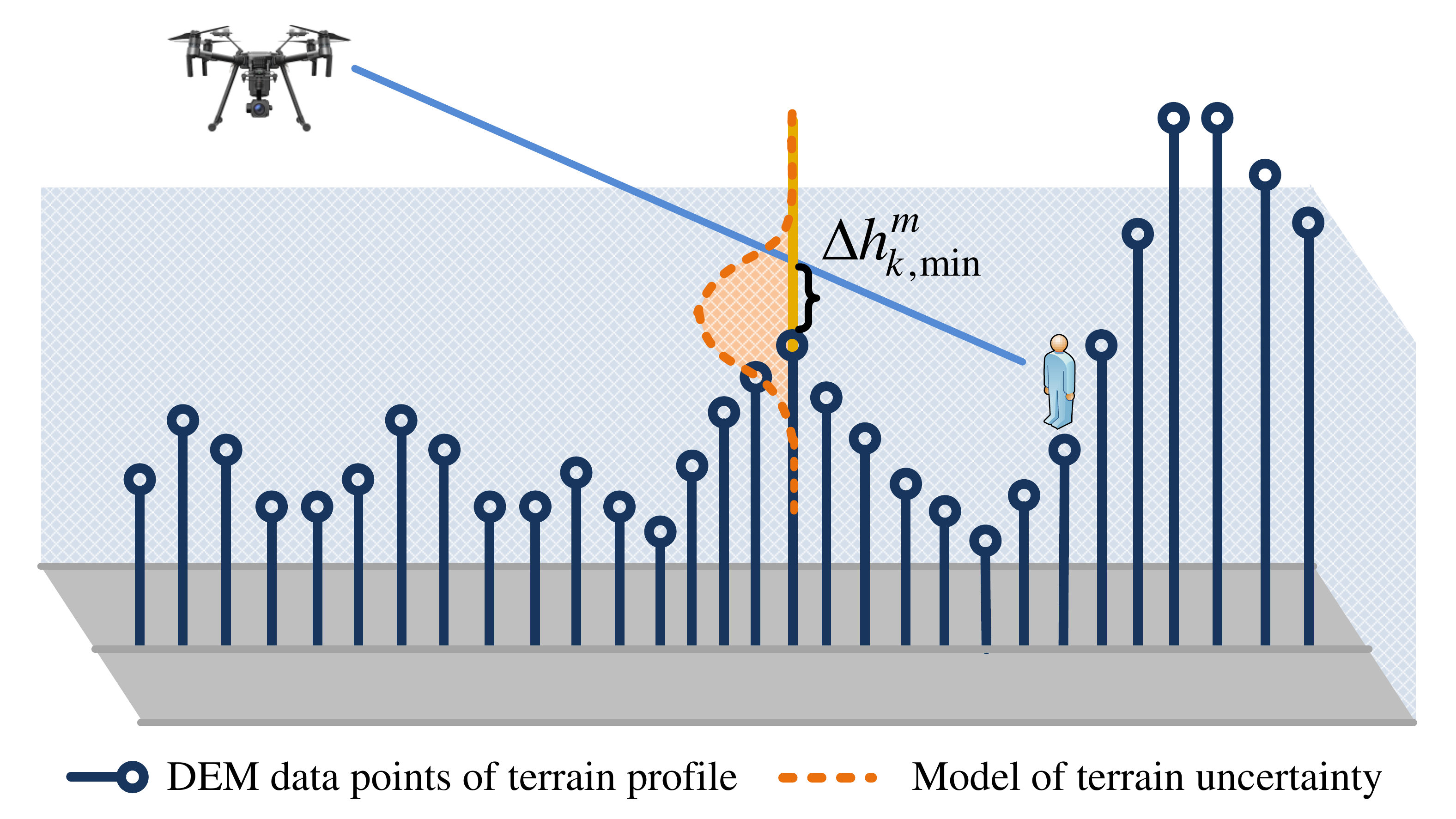}
\caption{Probability model of signal propagation condition.}
\label{fig_7}
\end{figure}
where $\sigma _h^2$ denotes the variance of terrain uncertainty. The prior probability of LoS condition is defined as the probability that $\Delta \tilde h_{k,\min }^m$ is positive, which can be expressed as
\begin{equation}
P_{k,LoS}^m \!=\! P\left( {\Delta h_{k,\min }^m \!+\! {n_h} \!>\! 0} \right) \!=\! 1 \!-\! {F_{{n_h}}}\left( { - \Delta h_{k,\min }^m} \right),
\end{equation}
where ${F_x}\!\left( t \right)\!=\!\frac{1}{{{\sigma _x}\sqrt {2\pi } }}\int_{ - \infty }^t {\exp \left( { - \frac{{{{\left( {x - {\mu _x}} \right)}^2}}}{{2\sigma _x^2}}} \right)dx}$ denotes the cumulative distribution function (CDF) of the random variable $x$, and ${\mu _x}$ is its mean value.

Contrary to the LoS condition, the prior probability that the LoS path does not exist is $P_{k,No\_LoS}^m = 1 - P_{k,LoS}^m$. In the absence of LoS path, two types of potential propagation conditions still need to be distinguished, namely the blockage condition and the NLoS condition. The basis for distinguishing these two conditions is whether the received reflected signals are strong enough to be detected by the receiver. In this article, we employ the log-distance path loss model with log-normal shadowing to calculate the strength and signal-to-noise (SNR) ratio of the reflected signals. In the absence of LoS path, the path loss between the k-th SP and the m-th sample point can be written as
\begin{equation}
PL_k^m\left( {dB} \right) = {\beta _0} + 10{\alpha _N}\log (l_k^m) + {\psi _N},
\end{equation}
where ${\beta _0} = 20\log \left( {\frac{{4\pi {f_c}}}{c}} \right)$ is the reference path loss at a distance of 1m; ${f_c}$ and $c$ are the main frequency of the A2G channel and the speed of light, respectively; ${\alpha _N}$ is the path loss exponent (PLE) when the LoS path does not exist; $l_k^m$\begin{scriptsize}$= \sqrt {{{\left( {{h_B} - h_U^m} \right)}^2} + {{\left\| {{{\bf{w}}_k} - {{\bf{u}}_m}} \right\|}^2}}$\end{scriptsize} is the 3-D distance between the SP and the sample point; ${\psi _N} \!\sim\! {\cal N}\left( {0,\sigma _N^2} \right)$ is a log-normal variable representing the shadowing, and $\sigma _N^2$ is its variance. In practice, the values of parameters ${\alpha _N}$ and ${\sigma _N^2}$ can be set or changed flexibly according to the local environment. As will be described in the next subsection, the proposed system uses TWR-based positioning to locate the ground user. The implementation of TWR requires the exchange of messages between two devices \cite{TWR_Intr_2}, which means that both the UAV platform and the user need to transmit and receive PRS signal. In practical applications, the transceiver carried by the UAV platform tends to have a larger transmit power than user's handheld device, which means that the signal transmitted by the latter is more difficult to detect. Therefore, the occurrence of NLoS condition depends directly on the SNR of the reflected signals transmitted by the user device and received at the UAV platform, which can be expressed as
\begin{equation}
SNR_{k,U \to B}^m\left( {dB} \right) = {P_{t,U}} - PL_k^m - {P_{{n_0}}},
\end{equation}
where ${P_{t,U}}$ denotes the constant transmit power of user device in dBm, and ${P_{{n_0}}}$ is the noise power in dBm. The reflected signals are considered to be detectable if the corresponding SNR exceeds a threshold $SN{R_{\min }}$, whose value is related to the sensitivity of the airborne transceiver. It is worth noting that although the distance between the SP and the sample point is known, the detection of reflected signals is still a random event due to the shadowing component ${\psi _N}$ in (3). In addition, we assume that the terrain uncertainty and the shadowing are independent of each other. Thus, the prior probability of NLoS condition is defined as the joint probability that the LoS path does not exist and $SNR_{k,U \to B}^m$ exceeds $SN{R_{\min }}$, which can be expressed as
\begin{equation}
\begin{split}
P_{k,NLoS}^m &\!=\! P\left( {SNR_{k,U \to B}^m \!>\! SN{R_{\min }},No\_LoS} \right)\\
&\!=\! P_{k,No\_LoS}^m \!\cdot\! P\!\left(\! {\left. {SNR_{k,U \to B}^m \!>\! SN{R_{\min }}} \!\right|\!No\_LoS} \!\right)\\
&\!=\! P_{k,No\_LoS}^m \!\cdot\! P\!\left( {{\psi _N} < {\psi _{\max }}} \right)\\
&\!=\! P_{k,No\_LoS}^m \!\cdot\! {F_{{\psi _N}}}\!\left( {{\psi _{\max }}} \right),
\end{split}
\end{equation}
where
\begin{equation}
{\psi _{\max }} = \left( {{P_{t,U}} \!-\! {P_{{n_0}}} \!-\! SN{R_{\min }}} \right) - \left( {{\beta _0} \!+\! 10{\alpha _N}\log \left( {l_k^m} \right)} \right).
\end{equation}

Moreover, the prior probability of blockage condition is given by
\begin{equation}
\begin{split}
P_{k,B}^m &= P\left( {SNR_{k,U \to B}^m \!\le\! SN{R_{\min }},No\_LoS} \right)\\
&= P_{k,No\_LoS}^m \cdot \left( {1 - {F_{{\psi _N}}}\left( {{\psi _{\max }}} \right)} \right)\\
&= 1 - P_{k,LoS}^m - P_{k,NLoS}^m.
\end{split}
\end{equation}

So far, the geometry-based probability model used in this article to characterize A2G channels in mountainous environments has been established. As long as the DEM of the local environment is available, the prior probability of each propagation condition corresponding to the A2G channel between the k-th SP and m-th sample point can be obtained utilizing equations (2), (5) and (7).

\subsection{Model of TWR-Based Positioning}
In the proposed system, TWR-based positioning is selected to support the positioning service. Compared with the OTDoA positioning widely used in terrestrial cellular networks, TWR-based positioning eases the constraint of time synchronization between anchor nodes, making it suitable for single-UAV systems \cite{TWR_Intr_1}. In TWR-based positioning, the time of flight (ToF) ${\tau _f}$ is obtained through an exchange of messages between the UAV platform and the ground user, as shown in Fig. 8(a). Specifically, the UAV platform first sends a ranging request message to the user device at local time $t_B^s$. Since the local clocks on both sides are not tightly synchronized in advance, the user device has an unknown clock bias $B$ with respect to the UAV platform at the beginning of the session. Moreover, the clock drift caused by the limited performance of internal oscillator exists in both local clocks, which affects the accuracy of the time interval measurement. The clock drifts of the UAV platform and the user device relative to the perfect clock are denoted by ${\delta _B}$ and ${\delta _U}$, respectively. After ${\tau _f} = {l \mathord{\left/ {\vphantom {l c}} \right. \kern-\nulldelimiterspace} c}$ seconds of propagation, the request message arrives at the user device and is detected at user's local time $t_U^r$, which can be expressed as
\begin{equation}
t_U^r = t_B^s + B + {\tau _f}\left( {1 + {\delta _U}} \right) + {e_U},
\end{equation}
where ${e_U}$ is the time of arrival (ToA) measurement error caused by the device's internal noise.
\begin{figure}[!t]
\centering
\includegraphics[height=1.48in,width=3.45in]{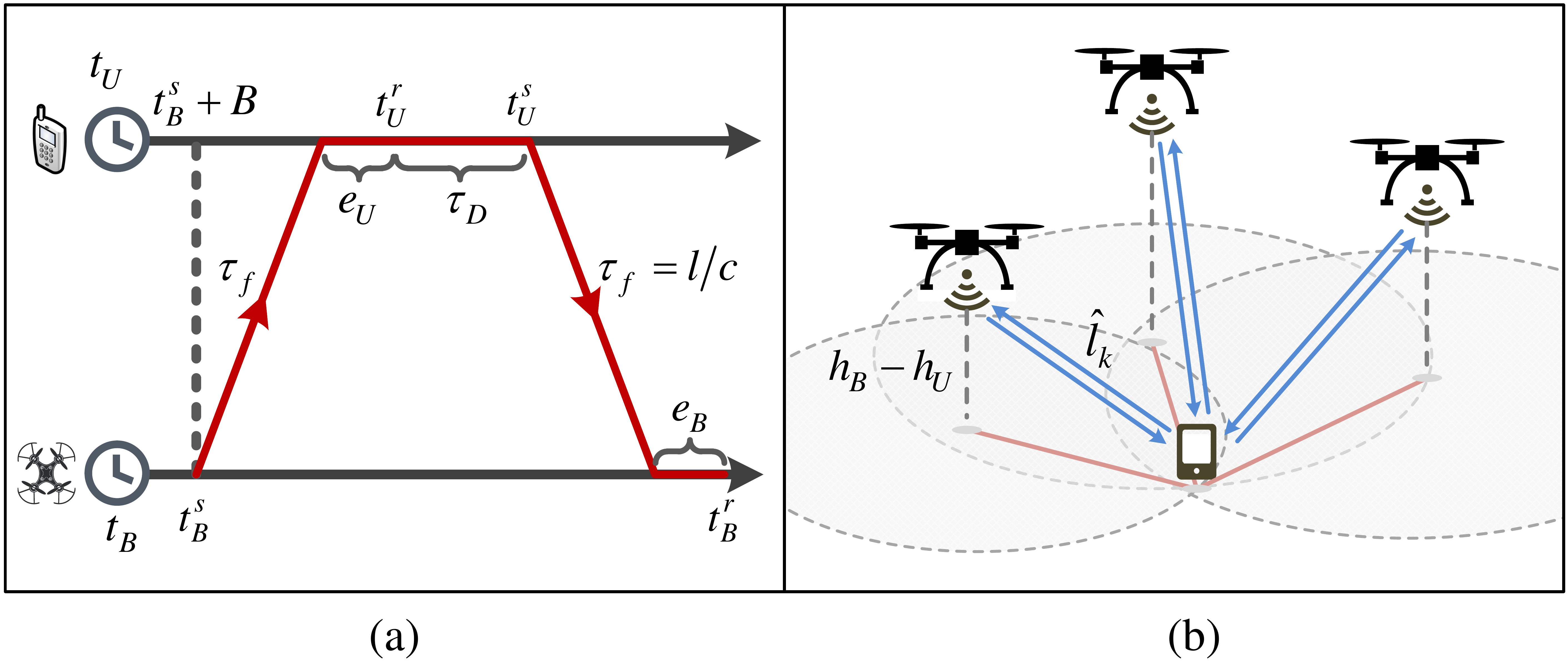}
\caption{Model of TWR-based positioning: (a) Time diagram of TWR and its (b) application example.}
\label{fig_8}
\end{figure}

After receiving the request message, the user device waits for a fixed time ${\tau _D}$ according to its own local clock, and then sends back a response message at local time
\begin{equation}
t_U^s = t_U^r + {\tau _D}.
\end{equation}
Finally, the response message is detected by the UAV platform at local time $t_B^r$, which can be expressed as
\begin{equation}
t_B^r = t_B^s + 2{\tau _f}\left( {1 \!+\! {\delta _B}} \right) + \left( {{e_U} \!+\! {\tau _D}} \right) \cdot \frac{{\left( {1 \!+\! {\delta _B}} \right)}}{{\left( {1 \!+\! {\delta _U}} \right)}} + {e_B}.
\end{equation}

After receiving the response message, the UAV platform can estimate the ToF according to its local clock, and the estimated value ${\hat \tau _f}$ is given by
\begin{equation}
\begin{split}
{{\hat \tau }_f} &= \frac{1}{2}\left[ {\left( {t_B^r - t_B^s} \right) - \left( {t_U^s - t_U^r} \right)} \right]\\
&= {\tau _f}\left( {1 \!+\! {\delta _B}} \right) + \frac{{{\tau _D}\left( {{\delta _B} \!-\! {\delta _U}} \right) \!+\! {e_U}\left( {1 \!+\! {\delta _B}} \right)}}{{2\left( {1 \!+\! {\delta _U}} \right)}} + \frac{{{e_B}}}{2}.\\
\end{split}
\end{equation}

The ToF estimation error can be expressed as
\begin{equation}
{\hat \tau _f} - {\tau _f} = {\tau _f}{\delta _B} + \frac{{{\tau _D}\left( {{\delta _B} \!-\! {\delta _U}} \right) \!+\! {e_U}\left( {1 \!+\! {\delta _B}} \right)}}{{2\left( {1 \!+\! {\delta _U}} \right)}} + \frac{{{e_B}}}{2}.
\end{equation}
Note that the ToF is typically much smaller than the response delay (${\tau _f} \ll {\tau _D}$), because the latter includes not only the turnaround time between the transmit and receive modes but also the packet duration of several milliseconds \cite{TWR_Intr_3}. Besides, we assume that the airborne transceiver is equipped with a high-performance oscillator. Thus, the clock drift of the UAV platform is negligible relative to that of the user's low-cost device, that is, ${\delta _B} \ll {\delta _U} \ll 1$. Then, the expression of ToF estimation error can be rewritten as
\begin{equation}
{\hat \tau _f} - {\tau _f} \approx  - \frac{{{\tau _D}{\delta _U}}}{2} + \frac{{{e_U}}}{2} + \frac{{{e_B}}}{2}.
\end{equation}

Under normal conditions where the NLoS propagation phenomenon and internal faults do not exist, the 3-D distance can be obtained from the estimated ToF utilizing the relationship $\hat l \!=\! {\hat \tau _f} \!\cdot\! c$. The corresponding ranging error can be expressed as
\begin{equation}
\Delta \hat l = c\left( {{{\hat \tau }_f} - {\tau _f}} \right) = {n_C} + {n_U} + {n_B},
\end{equation}
where ${n_C} =  - {{\left( {c \cdot {\tau _D}{\delta _U}} \right)} \mathord{\left/ {\vphantom {{\left( {c \cdot {\tau _D}{\delta _U}} \right)} 2}} \right. \kern-\nulldelimiterspace} 2}$ denotes the clock drift error; ${n_U} = {{\left( {c \cdot {e_U}} \right)} \mathord{\left/ {\vphantom {{\left( {c \cdot {e_U}} \right)} 2}} \right.
 \kern-\nulldelimiterspace} 2}$ and ${n_B} = {{\left( {c \cdot {e_B}} \right)} \mathord{\left/ {\vphantom {{\left( {c \cdot {e_B}} \right)} 2}} \right. \kern-\nulldelimiterspace} 2}$ are ranging errors caused by internal noise of the user device and the UAV platform, respectively. Without loss of generality, we model the clock drift error as a noise component, which follows a zero-mean Gaussian distribution:
\begin{equation}
{n_C} \sim {\cal N}\left( {0,\sigma _C^2} \right),
\end{equation}
where $\sigma _C^2$ denotes the variance of the clock drift error. If the crystal tolerance of user's oscillator ${O_U}$ is known, the value of parameter $\sigma _C^2$ can be determined using the following equation:
\begin{equation}
\sigma _C^2 = {\left( {\frac{{c \cdot {\tau _D}}}{2}} \right)^2} \!\cdot\! {\left( {\frac{{{O_U}}}{3}} \right)^2} = {\left(\! {\frac{{c \cdot {\tau _D}{O_U}}}{6}} \!\right)^2}.
\end{equation}
The other two terms on the right-hand side of equation (14) are also Gaussian noise components, that is, ${n_U} \sim {\cal N}\left( {0,\sigma _U^2} \right)$ and ${n_B} \sim {\cal N}\left( {0,\sigma _B^2} \right)$. In practice, the influence of devices' internal noise on ranging performance is typically negligible relative to the clock drift error \cite{TWR_Intr_2}, that is, $\sigma _U^2 \ll \sigma _C^2$, $\sigma _B^2 \ll \sigma _C^2$. Therefore, the expression of range measurement noise for TWR can be approximated as
\begin{equation}
\Delta \hat l = {n_C} + {n_U} + {n_B} \approx {n_C}.
\end{equation}

From the above analysis, it can be seen that no matter under which conditions, the range measurement noise of TWR depends mainly on the user's clock drift, whose variance is known and remains constant during the mission. As will be explained in the next section, this characteristic is very important for the detection of service failures.

As shown in Fig. 8(b), after the UAV platform traverses all SPs and performs TWR, the ground user's location can be estimated by solving the following equations:
\begin{equation}
{\hat l_k} = \sqrt {{{\left( {{h_B} \!-\! {h_U}} \right)}^2} + {{\left\| {{{\bf{w}}_k} \!-\! {\bf{u}}} \right\|}^2}}  + {n_C},\quad{\rm{  }}k \in {\cal K},
\end{equation}
where ${\hat l_k}$ is the range measurement obtained at the k-th SP; ${\bf{u}}$ and ${h_U}$ denote the user's horizontal coordinate and altitude, respectively. As mentioned at the beginning of this section, each user device is equipped with a barometer. During the TWR process, the accurate altitude measured by the barometer would be contained in the response message and sent to the UAV platform. Thus, only the horizontal coordinate ${\bf{u}}$ needs to be estimated, that is to say, the localization problem studied in this article is still a 2-D localization problem. In addition, when predicting the service reliability, the altitude of each sample point $h_U^m$ is set as the sum of the corresponding DEM data and the average height of handheld devices (1.5m).

\subsection{Model of Service Failure}
In the proposed system, the potential conditions in the range measurement between the k-th SP and m-th sample point can be divided into three categories: 1) blockage condition (B), 2) failure condition (F) and 3) normal condition (N). As analyzed in subsection A, the prior probability of blockage condition is $P_{k,B}^m$. On the contrary, the prior probability that the range measurement can be obtained (O) is $P_{k,O}^m = 1 - P_{k,B}^m$. In this article, the NLoS propagation and internal faults are considered as the two major causes of service failures, and are independent of each other. The prior probability of internal fault ${P_{IF}}$ could be determined in advance based on the reliability of the airborne transceiver and user device, and will not change during the positioning process. Thus, the prior probability of failure condition is given by
\begin{equation}
P_{k,F}^m = P_{k,LoS}^m{P_{IF}} + P_{k,NLoS}^m.
\end{equation}
The prior probability of normal condition where neither NLoS propagation nor internal fault occurs can be expressed as
\begin{equation}
P_{k,N}^m = P_{k,Los}^m \cdot \left( {1 - {P_{IF}}} \right).
\end{equation}
Moreover, the conditional probabilities of failure and normal conditions in the presence of range measurement are denoted as $P_{\left.\! {k,F} \!\right|O}^m \!=\! {{P_{k,F}^m} \mathord{\left/ {\vphantom {{P_{k,F}^m} {P_{k,O}^m}}} \right. \!\kern-\nulldelimiterspace\!} {P_{k,O}^m}}$ and $P_{\left.\! {k,N} \!\right|O}^m \!=\! {{P_{k,N}^m} \mathord{\left/ {\vphantom {{P_{k,N}^m} {P_{k,O}^m}}} \right. \!\kern-\nulldelimiterspace\!} {P_{k,O}^m}}$, respectively.

The influence of NLoS propagation and internal faults on TWR can be modeled as significant biases introduced in range measurements. Thus, the expression of range measurement under failure conditions can be written as
\begin{equation}
{\hat l_k} = \sqrt {{{\left( {{h_B} \!-\! {h_U}} \right)}^2} \!+\! {{\left\| {{{\bf{w}}_k} \!-\! {\bf{u}}} \right\|}^2}}  + {n_C} + {b_k},\;\;{\rm{  }}k \in {\cal K}.
\end{equation}
It is worth noting that if the measurement bias ${b_k}$ is caused only by NLoS propagation, its value is always positive. Otherwise, the value of ${b_k}$ could be either positive or negative. In practice, the absolute value of the bias ${b_k}$ is typically much larger than that of the measurement noise ${n_C}$. Although it is possible that the positive bias of NLoS propagation and the negative bias caused by internal faults cancel each other out exactly, the probability of such events is extremely small and will not be considered in the following.

\section{Reliability Prediction for UAV-Enabled Positioning}
There are many mature techniques that can be used in the proposed system to detect failures. Besides, during the positioning process, the reliability (i.e., the minimum detectable error) of failure detection could be evaluated in real time utilizing the methods presented in existing research \cite{Multi_Fault_1}. Thus, the focus of this section is not online fault detection or reliability evaluation, but the offline reliability prediction before the UAV takes off. For the proposed system, once the UAV is launched without adequate reliability prediction, the reliability of failure detection and the accuracy of positioning results are very likely to fail to meet the mission requirements, resulting in a waste of time and resources (fuel or battery). Therefore, it is necessary to perform reliability prediction before the mission starts. However, compared with the online reliability evaluation, the realization of the offline reliability prediction is much more difficult. The major challenge is that due to the uncertainty of user's location and probabilistic propagation condition, it is impossible to determine whether the range measurement corresponding to a certain SP could be obtained or not (i.e., the observation condition) before the TWR is performed.

This section presents our proposed reliability prediction method. The failure detection method used in this article is introduced in subsection A. Then, in subsection B, we divide all possible events in the positioning process into multiple ``observation events'' and ``failure events'' according to the visibility and status (normal or failure) of each SP, and the prior probability of each event is derived. Finally, the reliability of the positioning service is predicted based on the minimum detectable error in each failure event.

In addition, we assume that the mission requires the proposed system to detect failures that cause position error larger than ${\eta _{REQ}}$ in either x- or y-direction. The tolerable false alarm (FA) rate and missed detection (MD) rate are denoted by $P_{FA}^{REQ}$ and $P_{MD}^{REQ}$, respectively.

\subsection{Failure Detection based on LS-Residuals}
In this article, the least-squares-residuals (LS) method, a well-known fault detection method originally developed for GNSS applications \cite{RAIM_Intr_1,Multi_Fault_1}, is applied to the proposed system to detect failures. We assume that during the mission, the UAV platform can obtain range measurements at $N$ SPs. Then, the basic measurement equation used in LS method is obtained through the linearization of equation (21), which can be expressed as
\begin{equation}
{{\bf{l}}_ * } = {{\bf{H}}_ * }{\bf{u}} + {{\bf{n}}_ * } + {{\bf{b}}_ * },
\end{equation}
where ${{\bf{l}}_ * } \!=\! {\left[ {{{\hat l}_1}, \!\cdots\! ,{{\hat l}_N}} \right]^T}$ is the $N \!\times\! 1$ measurement vector, and ${\hat l_k}$ represents the range measurement corresponding to the k-th SP; ${\bf{u}} = {\left[\! {{x_U},{y_U}} \!\right]^T}$ is the $2 \!\times\! 1$ state vector, representing the unknown horizontal coordinate of the ground user; ${{\bf{H}}_ * } = {\left[\! {{{\left(\! {{{\partial {{\hat l}_1}}\! \mathord{\left/
 {\vphantom {{\partial {{\hat l}_1}} {\!\partial {\bf{u}}}}} \right.
 \kern-\nulldelimiterspace} {\!\partial {\bf{u}}}}} \!\right)}^T}\!, \!\cdots\! ,\!{{\left(\! {{{\partial {{\hat l}_N}} \!\mathord{\left/
 {\vphantom {{\partial {{\hat l}_N}} {\!\partial {\bf{u}}}}} \right.
 \kern-\nulldelimiterspace} {\!\partial {\bf{u}}}}} \!\right)}^T}} \!\right]^T}$ is the $N \!\times\! 2$ Jacobian matrix of ranging equations; ${{\bf{n}}_ * }$ is the $N \!\times\! 1$ vector of zero-mean Gaussian measurement noise due to the clock drift, and its covariance matrix is denoted by ${{\bf{V}}_ * } \!=\! \sigma _C^2 \cdot {{\bf{I}}_N}$; ${{\bf{b}}_ * }$ is the $N \!\times\! 1$ fault vector composed of measurement biases caused by faults, and ${{\bf{b}}_ * } = {{\bf{0}}_{N \times 1}}$ under normal conditions. For the sake of convenience, we normalize the above equation with the covariance matrix ${{\bf{V}}_ * }$ \cite{Multi_Fault_1}. The normalized measurement equation is given by
\begin{equation}
{\bf{l}} = {\bf{Hu}} + {\bf{n}} + {\bf{b}},
\end{equation}
where ${\bf{I}} \!=\! {\bf{V}}_ * ^{ - {1 \mathord{\left/ {\vphantom {1 2}} \right. \kern-\nulldelimiterspace} 2}}{{\bf{I}}_ * }$, ${\bf{H}} \!=\! {\bf{V}}_ * ^{ - {1 \mathord{\left/ {\vphantom {1 2}} \right. \kern-\nulldelimiterspace} 2}}{{\bf{H}}_ * }$, ${\bf{n}} \!=\! {\bf{V}}_ * ^{ - {1 \mathord{\left/ {\vphantom {1 2}} \right. \kern-\nulldelimiterspace} 2}}{{\bf{n}}_ * }$ and ${\bf{b}} \!=\! {\bf{V}}_ * ^{ - {1 \mathord{\left/ {\vphantom {1 2}} \right. \kern-\nulldelimiterspace} 2}}{{\bf{b}}_ * }$. After normalization, the covariance matrix (${\bf{V}}$) of the noise vector ${\bf{n}}$ becomes an $N \!\times\! N$ identity matrix, that is, ${\bf{n}} \!\sim\! {\cal N}\left( {{{\bf{0}}_{N \!\times\! 1}},{{\bf{I}}_N}} \right)$. Then, the least-squares solution of the normalized equation can be expressed as
\begin{equation}
{\bf{\hat u}} = {\left( {{{\bf{H}}^T}{\bf{H}}} \right)^{ - 1}}{{\bf{H}}^T}{\bf{l}} = {\bf{Gl}}.
\end{equation}
Specifically, the estimated x- and y-coordinates can be written as \cite{Multi_Fault_1}
\begin{equation}
{\hat x_U} = {\bm{\alpha }}_x^T{\bf{\hat u}} = {\bf{s}}_x^T{\bf{l}},
\end{equation}
\begin{equation}
{\hat y_U} = {\bm{\alpha }}_y^T{\bf{\hat u}} = {\bf{s}}_y^T{\bf{l}},
\end{equation}
where ${{\bm{\alpha }}_x} \!=\! {\left[ {1,0} \right]^T}$, ${{\bm{\alpha }}_y} \!=\! {\left[ {0,1} \right]^T}$, ${\bf{s}}_x^T \!=\! {\bm{\alpha }}_x^T{\bf{G}}$ and ${\bf{s}}_y^T \!=\! {\bm{\alpha }}_y^T{\bf{G}}$.
\begin{figure}[!t]
\centering
\includegraphics[height=1.60in,width=3.05in]{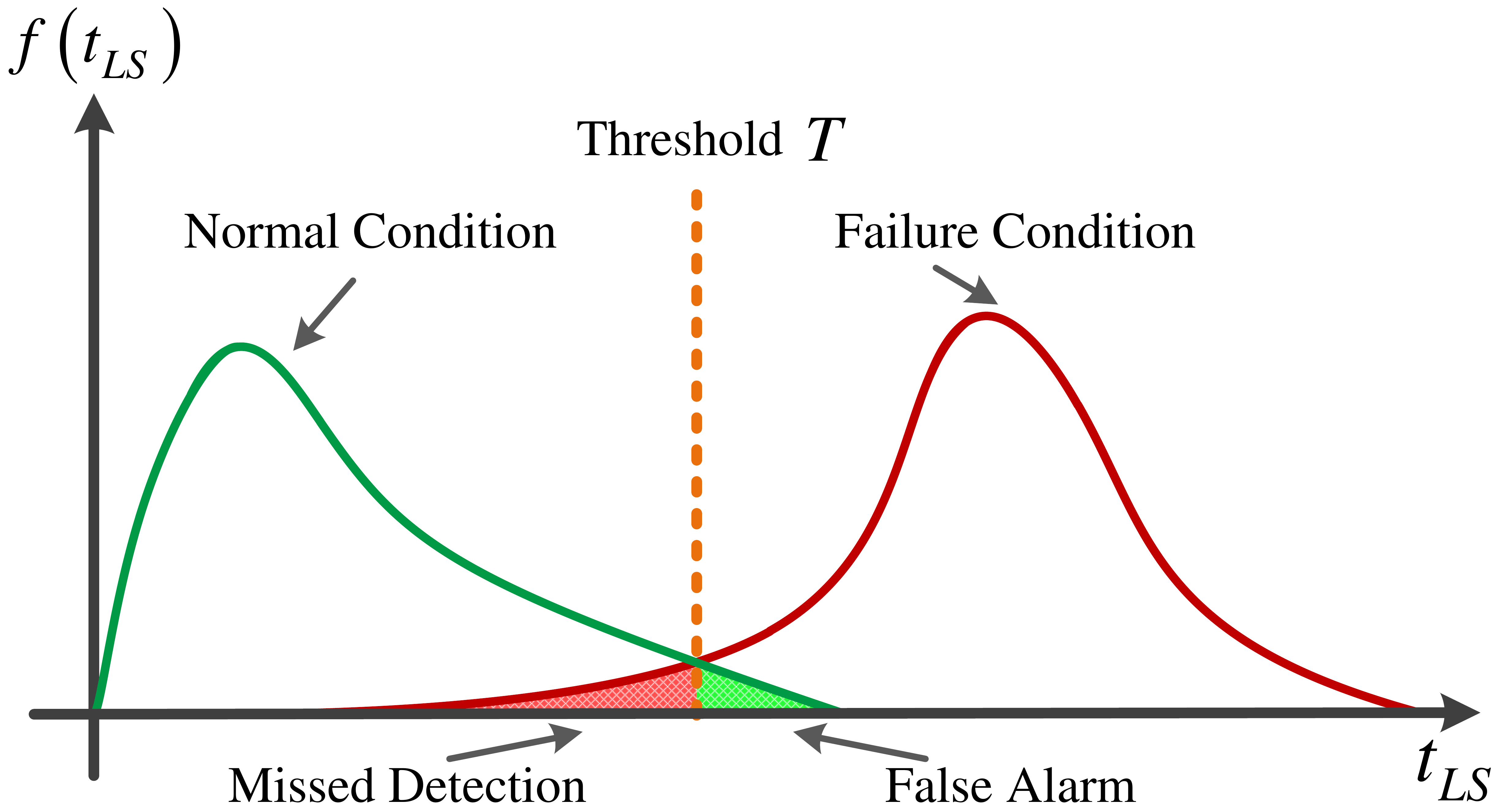}
\caption{Distributions of the test statistic.}
\label{fig_9)}
\end{figure}

The expression and distribution of the estimate error are given by
\begin{equation}
{\bm{\varepsilon }} = {\bf{\hat u}} \!-\! {\bf{u}} = {\bf{G}}\left( {{\bf{Hu}} \!+\! {\bf{n}} \!+\! {\bf{b}}} \right) \!-\! {\bf{u}} = {\bf{G}}\left( {{\bf{n}} \!+\! {\bf{b}}} \right),
\end{equation}
\begin{equation}
{\varepsilon _x} \!=\! {\bm{\alpha }}_x^T{\bm{\varepsilon }} \!=\! {\bf{s}}_x^T\left( {{\bf{n}} + {\bf{b}}} \right),\quad{\varepsilon _y} \!=\! {\bm{\alpha }}_y^T{\bm{\varepsilon }} \!=\! {\bf{s}}_y^T\left( {{\bf{n}} + {\bf{b}}} \right),
\end{equation}
\begin{equation}
{\varepsilon _x} \!\sim\! \left\{\! {\begin{array}{*{20}{c}}
{{\cal N}\left( {0,{\bf{s}}_x^T{{\bf{s}}_x}} \right),\quad\;\;\ {\rm{Normal\ condition,}}}\\
{{\cal N}\left( {{\bf{s}}_x^T{\bf{b}},{\bf{s}}_x^T{{\bf{s}}_x}} \right),\;\ {\rm{Failure\ condition}}{\rm{.}}}
\end{array}} \right.
\end{equation}
It can be clearly seen from the above equations that faults do introduce a large bias in the positioning result, making the location estimator no longer unbiased. To address this problem, the LS method performs the failure detection based on the differences (residuals) between the actual measured ranges and the predicted ranges, which can be expressed as
\begin{equation}
{\bf{\hat l}} = {\bf{H\hat u}} = {\bf{HGl}},
\end{equation}
\begin{equation}
{\bf{r}} = {\bf{l}} - {\bf{\hat l}} = \left( {{{\bf{I}}_N} \!-\! {\bf{HG}}} \right){\bf{l}} = \left( {{{\bf{I}}_N} \!-\! {\bf{HG}}} \right)\left( {{\bf{n}} \!+\! {\bf{b}}} \right),
\end{equation}
where ${\bf{\hat l}}$ is the $N \times 1$ vector of predicted ranges based on the estimated location, and ${\bf{r}}$ is the residual vector. The squared norm of the vector ${\bf{r}}$ is employed as the test statistic for failure detection, which can be written as
\begin{equation}
{t_{LS}} \buildrel \Delta \over = {\left\| {\bf{r}} \right\|^2} = {{\bf{r}}^T}{\bf{r}}.
\end{equation}
\begin{figure*}[htbp]
\centering
\includegraphics[height=2.30in,width=5.50in]{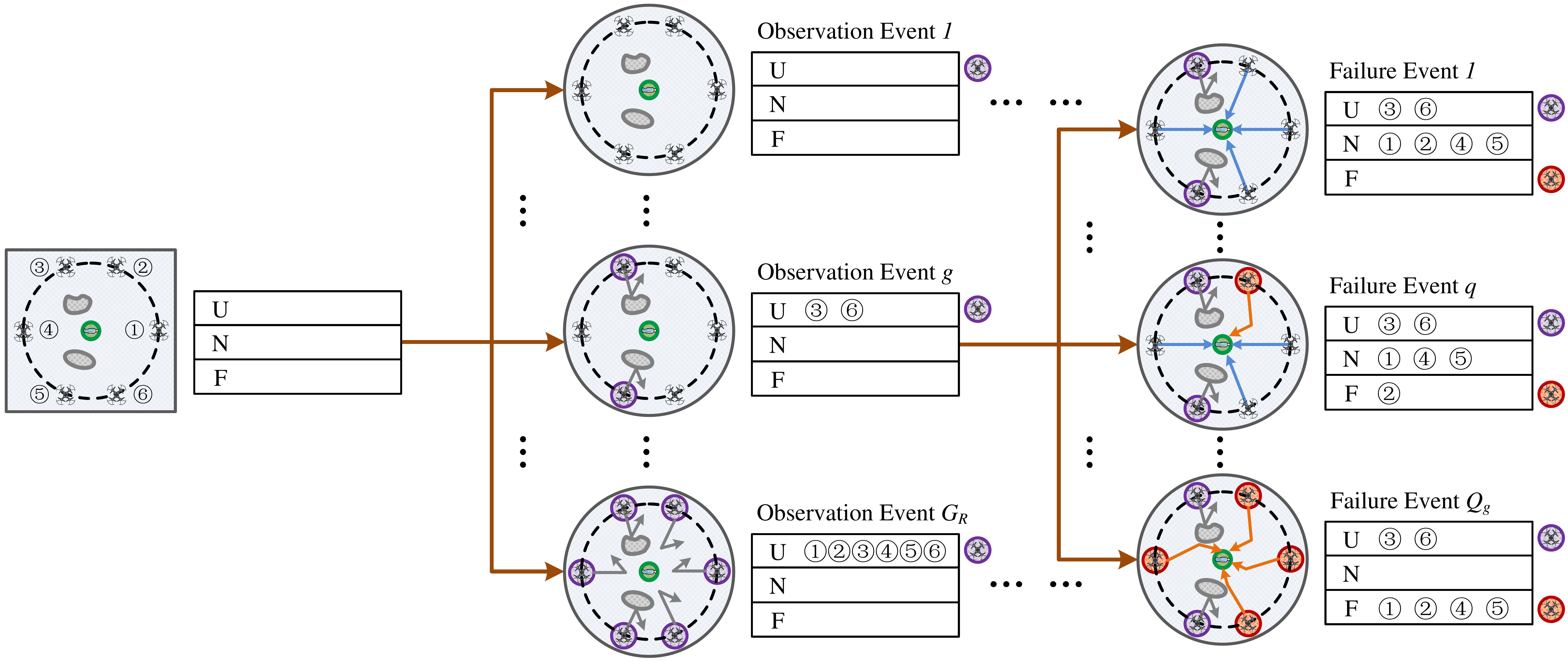}
\caption{Event tree.}
\label{fig_10}
\end{figure*}
Under normal conditions, the test statistic ${t_{LS}}$ is the sum of the squares of multiple Gaussian random variables, which are obtained through a linear transformation of another set of independent zero-mean Gaussian variables with the same variance. Thus, ${t_{LS}}$ follows a chi-square distribution (green solid line in Fig. 9) with $N - 3$ degrees of freedom (DOF). On the contrary, ${t_{LS}}$ has a non-central chi-square distribution (red solid line) under failure conditions. The distributions of test statistic ${t_{LS}}$ under normal and failure conditions can be expressed as follows:
\begin{equation}
{t_{LS}} \!\sim\! \left\{\! {\begin{array}{*{20}{c}}\!
\!{{\chi ^2}\!\left(\! {N \!-\! 3} \!\right),\qquad\qquad\qquad\;\ {\rm{Normal\ condition,}}}\\
\!{{\chi ^2}\!\left(\! {N \!-\! 3,{{\bf{b}}^T}\!\left(\! {{\bf{I}} \!-\! {\bf{HG}}} \!\right)\!{\bf{b}}} \!\right),\;{\rm{Failure\ condition}}{\rm{.}}}
\end{array}} \right.
\end{equation}

As shown in Fig. 9, if we set a decision threshold $T$ for the test statistic ${t_{LS}}$, the failure detection can be realized with the following decision rule:

\centerline{\emph{If: ${t_{LS}} \ge T$, declare ``service failure.'';}}

\centerline{\;\,\emph{If: ${t_{LS}} < T$, declare ``service succeed.''.}}

\subsection{Observation Events and Failure Events}
Before the mission starts, the visibility of SPs is unknown. In other words, we cannot be sure whether the UAV platform can obtain the desired range measurement at a certain SP. However, there is no doubt that the number and geometry of available SPs will affect the reliability of failure detection. Therefore, every possible observation event must be taken into consideration when making reliability prediction, as shown in Fig. 10.

Assume that the system has set $K$ SPs for the UAV, then the number of possible observation events is $G = {2^K}$. Among all observation events, there are ${G_{SU}} \!=$\begin{scriptsize}$\left(\! {\begin{array}{*{20}{c}}K\\2\end{array}} \!\right) + \left(\! {\begin{array}{*{20}{c}}K\\1\end{array}} \!\right) + 1$\end{scriptsize} ``service-unavailable (SU)'' events in which the available SPs are too few to support positioning service. Moreover, in ${G_{PO}} \!=$\begin{scriptsize}$\left(\! {\begin{array}{*{20}{c}}K\\3\end{array}} \!\right)$\end{scriptsize} ``positioning-only (PO)'' observation events, the number of available SPs is just enough for positioning service but still not enough for failure detection. Excluding the above two types of events, the remaining ${G_R} = G - {G_{SU}} - {G_{PO}}$ observation events in which the failure detection function can be implemented are represented by the set
\begin{equation}
{\cal O} = \left\{ {{{\bf{o}}_1},{{\bf{o}}_2}, \cdots ,{{\bf{o}}_{{G_R}}}} \right\},
\end{equation}
where the subset
\begin{equation}
{{\bf{o}}_g} = \left\{ {a_1^g, \cdots ,a_{{A_g}}^g,u_1^g, \cdots ,u_{{U_g}}^g} \right\}
\end{equation}
denotes the g-th observation event with ${A_g}$ available SPs and ${U_g}$ unavailable SPs (${A_g} + {U_g} = K$); $a_i^g$ and $u_j^g$ are the index numbers of the i-th available SP and j-th unavailable SP in observation event ${{\bf{o}}_g}$, respectively. For the m-th sample point of user’s location, the prior probability of the g-th observation event is given by
\begin{equation}
P_g^m = \prod\limits_{i = 1}^{{A_g}} {P_{a_i^g,O}^m}  \cdot \prod\limits_{j = 1}^{{U_g}} {P_{u_j^g,B}^m} ,
\end{equation}
where the expressions of $P_{a_i^g,O}^m$ and $P_{u_j^g,B}^m$ are derived in Section II. Moreover, the conditional FA rate corresponding to the g-th observation event can be expressed as
\begin{equation}
\begin{split}
P_{\left. {FA} \right|{{\bf{o}}_g}}^m &= P\left( {\left. {\left| {{\varepsilon _x}} \right| \!<\! \eta  \cap \left| {{\varepsilon _y}} \right| \!<\! \eta  \cap {t_{LS}} \!\ge\! {T_g}} \right|{{\bf{o}}_g},{\bf{b}} \!=\! {\bf{0}}} \right)\\
&< P\left( {\left. {{t_{LS}} \ge {T_g}} \right|{{\bf{o}}_g},{\bf{b}} = {{\bf{0}}_{{A_g} \times 1}}} \right)\\
&= \int\limits_{{T_g}}^{ + \infty } {\chi _t^2\left( {{A_g} - 3,0} \right)} dt,
\end{split}
\end{equation}
where ${T_g}$ is the decision threshold for failure detection in observation event ${{\bf{o}}_g}$.

As analyzed in Section II.C, each range measurement obtained in the positioning process may be contaminated by failure. As shown in Fig. 10, according to the status of each SP, the observation event ${{\bf{o}}_g}$ can be further subdivided into one normal event in which no failure occurs and ${Q_g} = {2^{{A_g}}} - 1$ failure events. The prior probability of the normal event ${\bf{f}}_0^g$ is given by
\begin{equation}
P_{g,0}^m = P_g^m \cdot P_{\left. 0 \right|g}^m = P_g^m \cdot \prod\limits_{i = 1}^{{A_g}} {P_{a_i^g,\left. N \right|O}^m} .
\end{equation}
All potential failure events in the g-th observation event can be represented by the set
\begin{equation}
{{\cal F}_g} = \left\{ {{\bf{f}}_1^g,{\bf{f}}_2^g, \cdots ,{\bf{f}}_{{Q_g}}^g} \right\},
\end{equation}
where the subset
\begin{equation}
{\bf{f}}_q^g = \left\{ {n_1^{g,q}, \cdots ,n_{N_q^g}^{g,q},f_1^{g,q}, \cdots ,f_{F_q^g}^{g,q}} \right\}
\end{equation}
denotes the q-th failure event in observation event ${{\bf{o}}_g}$, in which there are $N_q^g$ normal SPs and $F_q^g$ failure SPs ($N_q^g + F_q^g = {A_g}$); $n_i^{g,q}$ and $f_j^{g,q}$ are the index numbers of the i-th normal SP and j-th failure SP in failure event ${\bf{f}}_q^g$, respectively. For the m-th sample point, the prior probability of the q-th failure event is given by
\begin{equation}
P_{g,q}^m = P_g^m \!\cdot\! P_{\left. q \right|g}^m = P_g^m \!\cdot\! \prod\limits_{i = 1}^{N_q^g} \!{P_{n_i^{g,q}\!,\!\left. N \right|O}^m}  \!\cdot\! \prod\limits_{j = 1}^{F_q^g} \!{P_{f_j^{g,q},\left. F \right|O}^m} ,
\end{equation}
where the expressions of conditional probabilities $P_{n_i^{g,q}\!,\!\left. N \right|O}^m$ and $P_{f_j^{g,q}\!,\!\left. F \right|O}^m$ can be found in Section II.C. It is noteworthy that for any failure event, there is a certain probability that the failure is not successfully detected, that is, the MD rate. The conditional MD rate corresponding to failure event ${\bf{f}}_q^g$ can be expressed as
\begin{equation}
\begin{split}
P_{\left. {MD} \right|{\bf{f}}_q^g}^m &= P\!\left(\! {\left. {\left( {\left| {{\varepsilon _x}} \right| \!\ge\! \eta  \cup \left| {{\varepsilon _y}} \right| \!\ge\! \eta } \right) \cap \left( {{t_{LS}} \!<\! {T_g}} \right)} \right|{\bf{f}}_q^g} \right)\\
&< P\left( {\left. {\left| {{\varepsilon _x}} \right| \ge \eta  \cap {t_{LS}} < {T_g}} \right|{\bf{f}}_q^g} \right)\\
&\quad\,+ P\left( {\left. {\left| {{\varepsilon _y}} \right| \ge \eta  \cap {t_{LS}} < {T_g}} \right|{\bf{f}}_q^g} \right).
\end{split}
\end{equation}

\subsection{Reliability Prediction and the Corresponding Metric}
\begin{algorithm}[!t]
\caption{Proposed Reliability Prediction Method}
\label{alg:Reliability Prediction}
\begin{algorithmic}[1]
\renewcommand{\algorithmicrequire}{ \textbf{Input:}}
\REQUIRE
DEM of realistic terrain, coordinate of each SP (${{\bf{w}}_k}$, $h_B$), number of sample points $M$, coordinate of each sample point (${{\bf{u}}_m}$, $h_U^m$), tolerable FA rate $P_{FA}^{REQ}$ and MD rate $P_{MD}^{REQ}$.
\renewcommand{\algorithmicrequire}{ \textbf{Initialization:}}
\REQUIRE ~\\
\STATE Utilize the DEM and the probability models introduced in Section II to calculate the prior probabilities ($P_{k,O}^m$, $P_{\left.\! {k,F} \!\right|O}^m$ and $P_{\left.\! {k,N} \!\right|O}^m$) of potential statuses of each SP according to parameters ${{\bf{w}}_k}$, $h_B$, ${{\bf{u}}_m}$ and $h_U^m$;\\
\renewcommand{\algorithmicrequire}{ \textbf{Event Classification:}}
\REQUIRE ~\\
\STATE Divide all possible events in positioning process into $G$ observation events according to the visibility of each SP, and further subdivided each of them into one normal event and ${Q_g}$ failure events according to SP's status; (34), (39);\\
\STATE Calculate the prior probability of each observation event ($P_g^m$), normal event ($P_{g,0}^m$) and failure event ($P_{g,q}^m$) based on $P_{k,O}^m$, $P_{\left.\! {k,F} \!\right|O}^m$ and $P_{\left.\! {k,N} \!\right|O}^m$; (36), (38) and (41);\\
\renewcommand{\algorithmicrequire}{ \textbf{Requirement Allocation:}}
\REQUIRE ~\\
\STATE Allocate the overall tolerable FA rate $P_{FA}^{REQ}$ among observation events with the scheme described in \textbf{\emph{Algorithm 2}}, and determine the threshold ${T_g}$ in each event ${{\bf{o}}_g}$;\\
\STATE Allocate the overall tolerable MD rate $P_{MD}^{REQ}$ among failure events and both directions (x and y) with the scheme described in \textbf{\emph{Algorithm 3}}, conditional MD rate $P_{\left. {MD} \right|{\bf{f}}_q^g}^{REQ,x}$ for the x-direction in event ${\bf{f}}_q^g$;\\
\renewcommand{\algorithmicrequire}{ \textbf{Reliability Prediction:}}
\REQUIRE
\FOR{$m=1$ to ${M}$}
\FOR{$g=1$ to ${G_R}$}
\FOR{$q=1$ to ${Q_g}$}
\STATE Calculate the minimum detectable errors in x- and y-directions ($\eta _{g,q}^{ * m,x}$ and $\eta _{g,q}^{ * m,y}$) based on ${T_g}$, $P_{\left. {MD} \right|{\bf{f}}_q^g}^{REQ,x}$ and $P_{\left. {MD} \right|{\bf{ f}}_q^g}^{REQ,y}$; (57)-(59);\\
\ENDFOR
\ENDFOR
\ENDFOR
\STATE Take ${\eta ^ * } = \mathop {\max }\limits_{m \in {\cal M}} \left\{ {\mathop {\max }\limits_{g \in {\cal G}} \left\{ {\mathop {\max }\limits_{q \in {{{\cal Q}}_g}} \left\{ {\eta _{g,q}^{ * m,x},\eta _{g,q}^{ * m,y}} \right\}} \right\}} \right\}$ as the overall reliability prediction results; (60)-(63);\\
\renewcommand{\algorithmicrequire}{ \textbf{Output:}}
\REQUIRE
${\eta ^ * }$.
\end{algorithmic}
\end{algorithm}
The proposed reliability prediction method is summarized in \emph{Algorithm 1}. It can be seen that the implementation of this method can be divided into four stages: ``initialization'', ``event classification'', ``requirement allocation'' and ``reliability prediction''. The operations in the first two stages have been introduced above. In this subsection, we mainly describe the technical details of the remaining two stages.

As mentioned in the previous subsection, for the proposed system, all possible events in the positioning process can be divided into several observation events, each of which can be further subdivided into multiple failure events. However, the mission requirement only gives the tolerable limits of overall FA and MD rates, and does not specify the tolerable conditional FA/MD rate corresponding to each observation event/failure event. As can be seen from equation (37), the conditional FA rate of each observation event ${{\bf{o}}_g}$ mainly depends on the value of the decision threshold ${T_g}$. From another perspective, if the tolerable conditional FA rate $P_{\left. {FA} \right|{{\bf{o}}_g}}^{REQ}$ is given, the feasible region of decision threshold can be determined according to the constraint $P_{\left. {FA} \right|{{\bf{o}}_g}}^m \le P_{\left. {FA} \right|{{\bf{o}}_g}}^{REQ}$. In practical applications, the decision threshold is typically set to the value that just satisfies the constraint, that is,
\begin{equation}
\int\limits_{{T_g}}^{ + \infty } {\chi _t^2\left( {{A_g} - 3,0} \right)} dt = P_{\left. {FA} \right|{{\bf{o}}_g}}^{REQ}.
\end{equation}
It can be clearly seen from the above equation that the allocation of tolerable FA rate among observation events determines the values of decision thresholds, thereby affecting the results of reliability prediction. Thus, in this section, we first introduce the FA rate allocation scheme (\emph{Algorithm 2}) adopted in the proposed reliability prediction method.

In the reliability prediction process, the overall FA rate is defined as the sum of the conditional FA rates of all possible observation events weighted by the prior probabilities of corresponding normal events, that is,
\begin{equation}
P_{FA}^m \buildrel \Delta \over = \sum\limits_{g = 1}^{{G_R}} {P_{\left. {FA} \right|{{\bf{o}}_g}}^m \cdot P_{g,0}^m}  + P_{FA,PO}^m,
\end{equation}
where $P_{FA,PO}^m$ is the FA rate of ${G_{PO}}$ ``positioning-only'' observation events. Once these events occur, the system will sound an alarm. Thus, the MD rate of these observation events is 0, and their FA rate can be approximated by their prior probability. In addition, for those “service unavailable” observation events, their FA and MD rates are both 0. Then, according to the mission requirement ($P_{FA}^m \le P_{FA}^{REQ}$) mentioned at the beginning of this section, the allocation of tolerable FA rate should satisfy the following constraint:
\begin{equation}
\sum\limits_{g = 1}^{{G_R}} {P_{\left. {FA} \right|{{\bf{o}}_g}}^{REQ} \cdot P_{g,0}^m}  \le P_{FA}^{REQ} - P_{FA,PO}^m.
\end{equation}
\begin{algorithm}[!t]
\caption{False Alarm (FA) Rate Allocation Scheme}
\label{alg:FA Rate Allication}
\begin{algorithmic}[1]
\renewcommand{\algorithmicrequire}{ \textbf{Input:}}
\REQUIRE
Number of observation events $G$, prior probability of the normal event in each observation event ($P_{g,0}^m$), overall tolerable FA rate $P_{FA}^{REQ}$.
\renewcommand{\algorithmicrequire}{ \textbf{Allocation:}}
\REQUIRE ~\\
\STATE Exclude ${G_{SU}} \!+\! {G_{PO}}$ observation events in which failure detection is unimplementable, and subtract their FA rates $P_{FA,PO}^m$ from overall tolerable FA rate $P_{FA}^{REQ}$; (45);\\
\STATE Sort the remaining ${G_R} = G \!-\! {G_{SU}} \!-\! {G_{PO}}$ observation events in ascending order of $P_{g,0}^m$, and find the first ${g^ * } - 1$ events with negligible FA rates, exclude them and subtract their FA rates $P_{FA,EXL}^{REQ}$ from $P_{FA} \!-\! P_{FA,PO}^m$; (46)-(49);\\
\STATE Allocate $P_{FA} \!-\! P_{FA,PO}^m \!-\! P_{FA,EXL}^{REQ}$ to the remaining ${G_R} \!-\! {g^ * } \!+\! 1$ observation events according to $P_{g,0}^m$; (50);\\
\STATE Determine the threshold ${T_g}$ in each observation event ${{\bf{o}}_g}$ based on the allocated conditional FA rate $P_{\left. {FA} \right|{{{\bf{o}}}_g}}^{REQ}$; (43);\\
\renewcommand{\algorithmicrequire}{ \textbf{Output:}}
\REQUIRE
$P_{\left. {FA} \right|{{{\bf{o}}}_g}}^{REQ}$ and ${T_g}$.
\end{algorithmic}
\end{algorithm}
When allocating the tolerable FA rate, we first sort the observation events in ascending order of the prior probabilities of corresponding normal events. The sorted observation events and corresponding prior probabilities are, respectively, denoted by sets
\begin{equation}
\bar {\cal O} = \left\{ {{{{\bf{\bar o}}}_1},{{{\bf{\bar o}}}_2}, \cdots ,{{{\bf{\bar o}}}_{{G_R}}}} \right\},
\end{equation}
\begin{equation}
\bar {\cal P}_0^m = \left\{ {\bar P_{1,0}^m,\bar P_{2,0}^m, \ldots ,\bar P_{{G_R},0}^m} \right\},
\end{equation}
and $\bar P_{g - 1,0}^m \le \bar P_{g,0}^m \le \bar P_{g + 1,0}^m$. It is worth noting that since the SPs are carefully selected, most of the observation events in set $\bar {\cal O}$ are very unlikely to occur, that is to say, the FA rates of these events are small and negligible. The index of the first observation event that needs to be considered in the FA rate allocation can be expressed as
\begin{equation}
{g^ * } = \min \left[ {\left. g \right|\sum\limits_{i = 1}^g {\bar P_{i,0}^m \ge P_{FA}^{REQ} \!-\! P_{FA,PO}^m} } \right].
\end{equation}
Then, the first ${g^ * } - 1$ observation events in set $\bar {\cal O}$ are excluded. Specifically, the tolerable conditional FA rates $P_{\left. {FA} \right|{{\bf{o}}_g}}^{REQ}$ ($1 \!\le\! g \!<\! {g^ * }$) of these observation events are set to 1, whereas their conditional MD rates are 0. After excluding these observation events, the constraint on FA rate allocation among the remaining events can be written as
\begin{equation}
\sum\limits_{g = {g^ * }}^{{G_R}} {P_{\left. {FA} \right|{{{\bf{\bar o}}}_g}}^{REQ} \!\cdot\! \bar P_{g,0}^m}  \le P_{FA}^{REQ} \!-\! P_{FA,PO}^m \!-\! P_{FA,EXL}^{REQ},
\end{equation}
where $P_{FA,EXL}^{REQ} \!=\! \sum\limits_{g = 1}^{{g^ * } - 1} {1 \!\cdot\! \bar P_{g,0}^m}$ denotes the tolerable FA rate of those excluded observation events. Finally, the remaining tolerable FA rate ($P_{FA}^{REQ} - P_{FA,PO}^m - P_{FA,EXL}^{REQ}$) is allocated according to the prior probabilities of the remaining normal events, and the FA rate allocated to event ${{\bf{\bar o}}_g}$ (${g^ * } \le g \le {G_R}$) is given by
\begin{equation}
\begin{split}
P_{FA,{{{\bf{\bar o}}}_g}}^{REQ} &= P_{\left. {FA} \right|{{{\bf{\bar o}}}_g}}^{REQ} \cdot \bar P_{g,0}^m\\
&= \frac{{\left(\! {P_{FA}^{REQ} \!-\! P_{FA,PO}^m \!-\! P_{FA,EXL}^{REQ}} \!\right) \!\cdot\! \bar P_{g,0}^m}}{{\sum\limits_{i = {g^ * }}^{{G_R}} {\bar P_{i,0}^m} }}.
\end{split}
\end{equation}
After the tolerable conditional FA rate of each observation event ($P_{\left. {FA} \right|{{{\bf{\bar o}}}_g}}^{REQ} = {{P_{FA,{{{\bf{\bar o}}}_g}}^{REQ}} \mathord{\left/
 {\vphantom {{P_{FA,{{{\bf{\bar o}}}_g}}^{REQ}} {\bar P_{g,0}^m}}} \right.
 \kern-\nulldelimiterspace} {\bar P_{g,0}^m}}$) is determined, the corresponding decision threshold ${T_g}$ can be obtained utilizing equation (43).

It can be seen from equation (42) that once the decision threshold is determined, the conditional MD rate of each failure event ${\bf{f}}_q^g$ depends on the tolerable limit $\eta$ of position error. From another point of view, for a certain tolerable conditional MD rate $P_{\left. {MD} \right|{\bf{f}}_q^g}^{REQ}$, there exists a set of values of parameter $\eta$ that satisfy the constraint $P_{\left. {MD} \right|{\bf{f}}_q^g}^m \le P_{\left. {MD} \right|{\bf{f}}_q^g}^{REQ}$, which is called ``detectable position error''. The minimum detectable error (${\eta ^ * }$) can effectively reflect the reliability of failure detection, making it extremely important in reliability prediction. In the following paragraphs, we introduce the MD rate allocation scheme (\emph{Algorithm 3}) adopted in the proposed reliability prediction method.
\begin{algorithm}[!t]
\caption{Missed Detection (MD) Rate Allocation Scheme}
\label{alg:MD Rate Allication}
\begin{algorithmic}[1]
\renewcommand{\algorithmicrequire}{ \textbf{Input:}}
\REQUIRE
Number of failure events in each observation event (${Q_g}$), prior probability of each observation event ($P_g^m$) and failure event ($P_{g,q}^m$), overall tolerable MD rate $P_{MD}^{REQ}$.
\renewcommand{\algorithmicrequire}{ \textbf{Allocation:}}
\REQUIRE
\STATE Allocate $P_{MD}^{REQ}$ to the remaining ${G_R} \!-\! {g^ * } \!+\! 1$ observation events according to $P_g^m$, ${P_{MD,{{{\bf{o}}}_g}}^{REQ}}$ to event ${{\bf{o}}_g}$; (52)-(53);\\
\FOR{$g={g^ * }$ to ${G_R}$}
\STATE Sort the ${Q_g}$ failure events in observation event ${{\bf{o}}_g}$ in ascending order of $P_{g,q}^m$ and find the first ${q^ * } \!-\! 1$ events with negligible MD rates, exclude them and subtract their MD rates $P_{MD,EXL,{{{\bf{o}}}_g}}^{REQ}$ from ${P_{MD,{{{\bf{o}}}_g}}^{REQ}}$; (54);\\
\STATE Allocate ${P_{MD,{{{\bf{o}}}_g}}^{REQ}} - P_{MD,EXL,{{{\bf{o}}}_g}}^{REQ}$ to the remaining ${Q_g} \!-\! {q^ * } \!+\! 1$ failure events according to $P_{g,q}^m$; (55);\\
\STATE Allocate the conditional MD rate ${P_{\left. {MD} \right|{\bf{f}}_q^g}^{REQ}}$ of failure event ${\bf{f}}_q^g$ equally to the x- and y-directions, $P_{\left. {MD} \right|{\bf{f}}_q^g}^{REQ,x}$ to the x-direction; (56);\\
\ENDFOR
\renewcommand{\algorithmicrequire}{ \textbf{Output:}}
\REQUIRE
$P_{\left. {MD} \right|{\bf{f}}_q^g}^{REQ,x}$ and $P_{\left. {MD} \right|{\bf{f}}_q^g}^{REQ,y}$.
\end{algorithmic}
\end{algorithm}

According to the mission requirement ($P_{MD}^m \le P_{MD}^{REQ}$), the allocation of tolerable MD rate should satisfy the following constraint:
\begin{equation}
\sum\limits_{g = 1}^{{G_R}} {P_{MD,{{{\bf{\bar o}}}_g}}^{REQ}}  = \sum\limits_{g = 1}^{{G_R}} {\sum\limits_{q = 1}^{{{\bar Q}_g}} {P_{\left. {MD} \right|{\bf{f}}_q^g}^{REQ} \!\cdot\! P_{g,q}^m} }  \le P_{MD}^{REQ},
\end{equation}
where the term $\sum\limits_{g = 1}^{{G_R}} {P_{MD,{{{\bf{\bar o}}}_g}}^{REQ}}$ means the allocation of tolerable MD rate among observation events, and the term $\sum\limits_{q = 1}^{{{\bar Q}_g}} {P_{\left. {MD} \right|{\bf{f}}_q^g}^{REQ} \!\cdot\! P_{g,q}^m}$ represents the allocation among multiple failure events in the same observation event. We first analyze the former. As mentioned above, only the last ${G_R} - {g^ * } + 1$ observation events in set $\bar {\cal O}$ need to be considered, while the tolerable conditional MD rates of the other ${g^ * } - 1$ events are set to 0. The MD rate allocated to event ${{\bf{\bar o}}_g}$ (${g^ * } \!\le\! g \!\le\! {G_R}$) is
\begin{equation}
P_{MD,{{{\bf{\bar o}}}_g}}^{REQ} = {{\left( {P_{MD}^{REQ} \cdot \bar P_{g,{\bf{f}}}^m} \right)} \mathord{\left/
 {\vphantom {{\left( {P_{MD}^{REQ} \!\cdot\! \bar P_{g,{\bf{f}}}^m} \right)} {\sum\limits_{i = {g^ * }}^{{G_R}} {\bar P_{i,{\bf{f}}}^m} }}} \right.
 \kern-\nulldelimiterspace} {\sum\limits_{i = {g^ * }}^{{G_R}} {\bar P_{i,{\bf{f}}}^m} }},
\end{equation}
where
\begin{equation}
\bar P_{g,{\bf{f}}}^m = \sum\limits_{q = 1}^{{{\bar Q}_g}} {\bar P_{g,q}^m}  = \bar P_g^m - \bar P_{g,0}^m
\end{equation}
is the sum of the prior probabilities of all possible failure events in observation event ${{\bf{\bar o}}_g}$; $\bar P_g^m$ and $\bar P_{g,q}^m$ denote the prior probabilities of observation event ${{\bf{\bar o}}_g}$ and the q-th failure event in it, respectively.

Then, we further consider the MD rate allocation among multiple failure events in the same observation event ${{\bf{\bar o}}_g}$. Similar to the operation in FA rate allocation, we first sort the ${\bar Q_g}$ failure events according to their prior probabilities, and then exclude those events with negligible prior probabilities. Let ${\bar {\cal F}_g} = \left\{ {{\bf{\bar f}}_1^g,{\bf{\bar f}}_2^g, \cdots ,{\bf{\bar f}}_{{{\bar Q}_g}}^g} \right\}$ be the set of sorted failure events, and ${q^ * }$ be the index of the first failure event that needs to be considered. The constraint on MD rate allocation among the remaining ${\bar Q_g} - {q^ * } + 1$ events can be written as
\begin{equation}
\sum\limits_{q = {q^ * }}^{{{\bar Q}_g}} {P_{\left. {MD} \right|{\bf{\bar f}}_q^g}^{REQ} \!\cdot\! \bar P_{g,q}^m}  \le P_{MD,{{{\bf{\bar o}}}_g}}^{REQ} \!-\! P_{MD,EXL,{{{\bf{\bar o}}}_g}}^{REQ},
\end{equation}
where $P_{MD,EXL,{{{\bf{\bar o}}}_g}}^{REQ} \!=\! \sum\limits_{q = 1}^{{q^ * } - 1} {1 \!\cdot\! \bar P_{g,q}^m}$ denotes the tolerable MD rate of those excluded failure events. Finally, the remaining tolerable MD rate ($P_{MD,{{{\bf{\bar o}}}_g}}^{REQ} \!-\! P_{MD,EXL,{{{\bf{\bar o}}}_g}}^{REQ}$) is allocated to each failure event ${\bf{\bar f}}_q^g$ (${q^ * } \le q \le {\bar Q_g}$) according to its prior probability:
\begin{equation}
\begin{split}
P_{MD,{\bf{\bar f}}_q^g}^{REQ} &= P_{\left. {MD} \right|{\bf{\bar f}}_q^g}^{REQ} \cdot \bar P_{g,q}^m \\
&= \frac{{\left( {P_{MD,{{{\bf{\bar o}}}_g}}^{REQ} - P_{MD,EXL,{{{\bf{\bar o}}}_g}}^{REQ}} \right) \cdot \bar P_{g,q}^m}}{{\sum\limits_{j = {q^ * }}^{{{\bar Q}_g}} {\bar P_{g,j}^m} }}.
\end{split}
\end{equation}

In addition, in equation (42), the conditional MD rate $P_{\left. {MD} \right|{\bf{f}}_q^g}^m$ is upper-bounded by the sum of the conditional MD rates in x- and y-directions. In the rest of this subsection, we will analyze the reliability of failure detection in x- and y-directions separately, and determine the minimum detectable error in each direction. Thus, the tolerable conditional MD rate ($P_{\left. {MD} \right|{\bf{\bar f}}_q^g}^{REQ}$) of failure event ${\bf{\bar f}}_q^g$ is further equally allocated to the two directions, that is,
\begin{equation}
P_{\left. {MD} \right|{\bf{\bar f}}_q^g}^{REQ,x} = P_{\left. {MD} \right|{\bf{\bar f}}_q^g}^{REQ,y} = {{P_{\left. {MD} \right|{\bf{\bar f}}_q^g}^{REQ}} \mathord{\left/
 {\vphantom {{P_{\left. {MD} \right|{\bf{\bar f}}_q^g}^{REQ}} 2}} \right.
 \kern-\nulldelimiterspace} 2},
\end{equation}
where $P_{\left. {MD} \right|{\bf{\bar f}}_q^g}^{REQ,x}$ and $P_{\left. {MD} \right|{\bf{\bar f}}_q^g}^{REQ,y}$ denote the conditional MD rates allocated to the x- and y-directions, respectively.

In the proposed system, the minimum detectable position error in all possible failure events and both directions is taken as the metric for reliability prediction. We first take the failure detection in x-direction as an example to analyze the minimum detectable error in failure event ${\bf{\bar f}}_q^g$. The key to determining the minimum detectable error is to find the worst fault vector that maximizes the position error while satisfying the constraint of tolerable conditional MD rate, that is,
\begin{align*}\tag{57}
&&&\mathop {\max }\limits_{{\bf{b}}_q^g}&& \left| {{\varepsilon _x}} \right|&\\
&&&\;\;\mbox{s.t.}&&P\left( {\left. {{t_{LS}} \!<\! {T_g}} \right|{\bf{b}}_{g,q}^{m,x}} \right) \le P_{\left. {MD} \right|{\bf{\bar f}}_q^g}^{REQ,x},&
\end{align*}
where ${\bf{b}}_{g,q}^{m,x}$ denotes the fault vector in failure event ${\bf{\bar f}}_q^g$. Noted that in the above equation, the conditional MD rate $P\left( {\left. {\left| {{\varepsilon _x}} \right| \ge \eta  \cap {t_{LS}} < {T_g}} \right|{\bf{b}}_{g,q}^{m,x}} \right)$ is replaced by the term $P\left( {\left. {{t_{LS}} < {T_g}} \right|{\bf{b}}_{g,q}^{m,x}} \right)$ since $P\left( {\left. {\left| {{\varepsilon _x}} \right| \ge \eta  \cap {t_{LS}} < {T_g}} \right|{\bf{b}}_{g,q}^{m,x}} \right) \le P\left( {\left. {{t_{LS}} < {T_g}} \right|{\bf{b}}_{g,q}^{m,x}} \right)$. As mentioned in \cite{Max_Slope}, the worst fault vector is also the one with the largest ``failure slope ($s$)''. The slope $s$ is defined as the ratio of the squared mean of the position error ${\varepsilon _x}$ over the non-centrality parameter of the test statistic ${t_{LS}}$, and can be written as
\begin{equation}\tag{58}
s \buildrel \Delta \over = \frac{{{{\bf{b}}^T}{{\bf{s}}_x}{\bf{s}}_x^T{\bf{b}}}}{{{{\bf{b}}^T}\left( {{\bf{I}} - {\bf{HG}}} \right){\bf{b}}}}.
\end{equation}

The methods for calculating the worst fault vector and the corresponding slope were described in \cite{Multi_Fault_1} and will not be repeated here. Let ${\bf{b}}_{g,q}^{ * m,x}$ be the worst fault vector, and $s_{g,q}^{ * m,x}$ be the largest failure slope. Then, the minimum detectable error in the x-direction in event ${\bf{\bar f}}_q^g$ is defined as the estimate bias cause by the worst fault vector, and can be expressed as
\begin{equation}\tag{59}
\eta _{g,q}^{ * m,x} \!\buildrel \Delta \over =\! \sqrt {s_{g,q}^{ * m,x} \!\cdot\! {{\left( {{\bf{b}}_{g,q}^{ * m,x}} \right)}^T}\left( {{\bf{I}} \!-\! {{\bf{H}}_g}{{\bf{G}}_g}} \right)\left( {{\bf{b}}_{g,q}^{ * m,x}} \right)} ,
\end{equation}
where ${{\bf{H}}_g}$ is the Jacobian matrix of the normalized ranging equations in observation event ${{\bf{\bar o}}_g}$, and ${{\bf{G}}_g} = {\left( {{\bf{H}}_g^T{{\bf{H}}_g}} \right)^{ - 1}}{\bf{H}}_g^T$. With the above operations, we can also determine the minimum detectable error $\eta _{g,q}^{ * m,y}$ in the y-direction.

For the m-th sample point of user's location, its corresponding reliability prediction result is defined as the largest of the minimum detectable errors in all possible failure events, which can be expressed as
\begin{equation}\tag{60}
{\eta ^{ * m}} \!= \mathop {\max }\limits_{g \in {\cal G}} \left\{ {\mathop {\max }\limits_{q \in {{\bar {\cal Q}}_g}} \left\{ {\eta _{g,q}^{ * m,x},\eta _{g,q}^{ * m,y}} \right\}} \right\},
\end{equation}
where ${\cal G} = \left\{ {{g^ * }, \cdots ,{G_R}} \right\}$, and ${\bar {\cal Q}_g} = \left\{ {{q^ * }, \cdots ,{{\bar Q}_g}} \right\}$. The prediction results in x- and y-directions can be written as
\begin{equation}\tag{61}
{\eta ^{ * m,x}} = \mathop {\max }\limits_{g \in {\cal G}} \left\{ {\mathop {\max }\limits_{q \in {{\bar {\cal Q}}_g}} \left\{ {\eta _{g,q}^{ * m,x}} \right\}} \right\},
\end{equation}
\begin{equation}\tag{62}
{\eta ^{ * m,y}} = \mathop {\max }\limits_{g \in {\cal G}} \left\{ {\mathop {\max }\limits_{q \in {{\bar {\cal Q}}_g}} \left\{ {\eta _{g,q}^{ * m,y}} \right\}} \right\}.
\end{equation}
Moreover, as mentioned at the beginning of Section II, the uncertainty area of user's location contains $M$ sample points, and the prior probability of the user being located at each sample point is equal. Thus, the worst predicted reliability in all sample points is taken as the prediction result of service reliability, that is
\begin{equation}\tag{63}
{\eta ^ * } = \mathop {\max }\limits_{m \in {\cal M}} \left\{ {{\eta ^{ * m}}} \right\} = \mathop {\max }\limits_{m \in {\cal M}} \left\{ {{\eta ^{ * m,x}},{\eta ^{ * m,y}}} \right\}.
\end{equation}

With the proposed reliability prediction method and the derived metric, system operators and mission commanders are capable of evaluating the reliability of positioning service before the mission starts, which is beneficial for decision-making. If the predicted detectable position error ${\eta ^ * }$ does not exceed the mission's tolerable limit ${\eta _{REQ}}$, the UAV is permitted to take off and provide service, because we are pretty sure that those failures that cause unacceptable position errors will be successfully detected in the positioning process and will not mislead decision-makers.

\section{Preliminary Study on Reliability Enhancement}
In mountainous environments with complex and highly variable terrain, the mission's requirements for service reliability are often difficult to meet, even for a set of carefully selected SPs. Fig. 11 shows the reliability prediction results obtained in a test scenario, where the tolerable limit of position error (${\eta _{REQ}}$) is set to 20m. It can be clearly seen that in the location uncertainty area, there are many sample points whose predicted detectable errors exceed ${\eta _{REQ}}$. If the user happens to be located at one of these sample points, failures that cause position errors larger than ${\eta _{REQ}}$ may not be effectively detected during the positioning process, making the positioning result unreliable. Therefore, it is necessary to improve the reliability prediction results by appropriately adjusting the locations of some SPs, that is, the reliability enhancement.
\begin{figure}[!t]
\centering
\includegraphics[height=3.15in,width=3.45in]{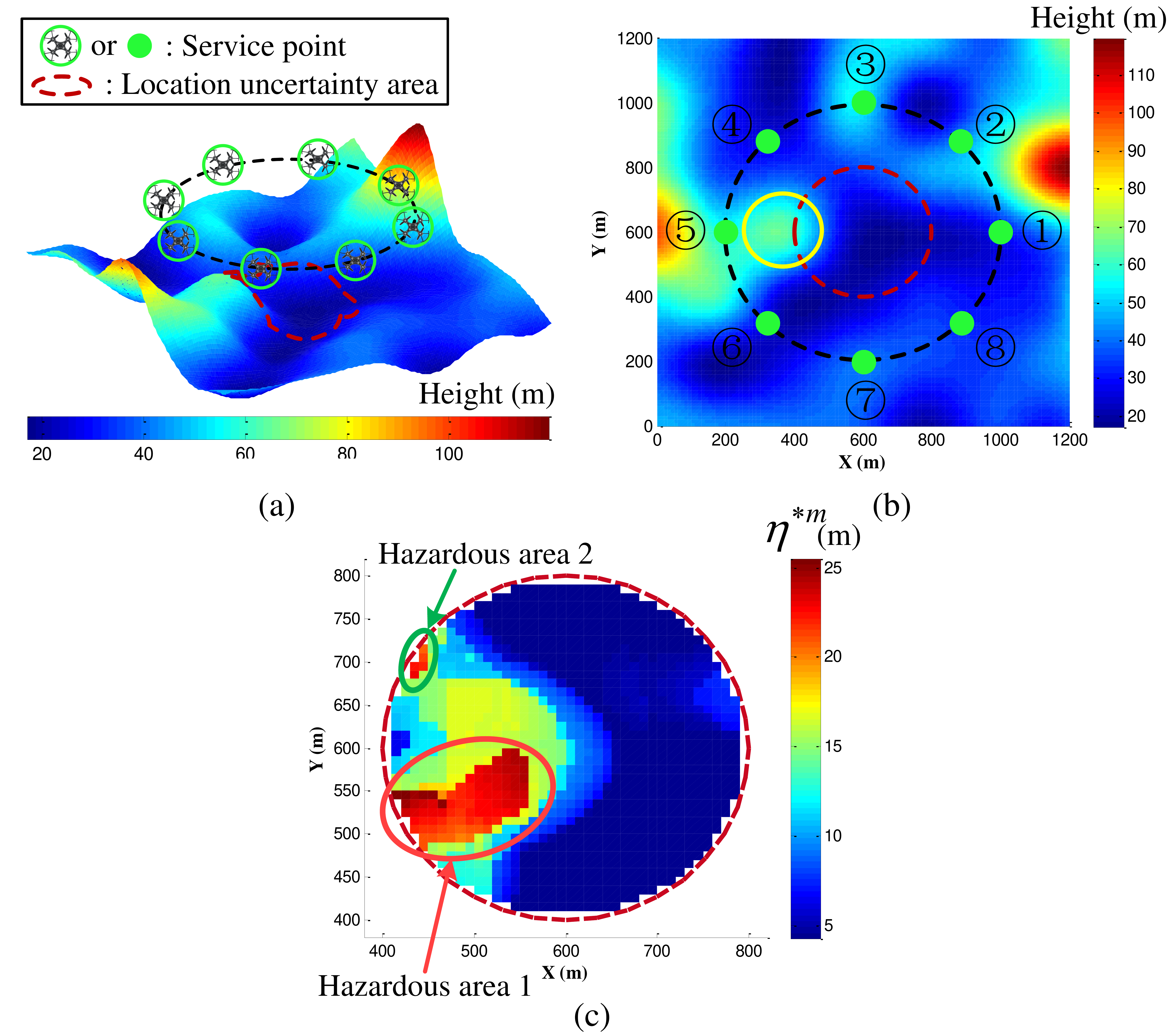}
\caption{Example of reliability prediction: (a) 3-D view and (b) top view of the test scenario, and (c) the corresponding prediction results (${\eta ^{ * m}}$).}
\label{fig_14)}
\end{figure}

The following two subsections introduce our preliminary study on reliability enhancement. In subsection A, we describe the data preprocessing procedures performed at the beginning of reliability enhancement. Subsection B presents our proposed voting-based method, which can be used to analyze the causes of unsatisfactory reliability and provide guidance for the adjustment of SPs.

\subsection{Identification and Segmentation of Hazardous Areas}
As can be seen from Fig. 11(c), sample points with unsatisfactory reliability are spatially clustered into two small blocks, which are called ``hazardous areas'' in this article. The aim of reliability enhancement is to analyze the causes of these areas and eliminate them based on the analysis results. Obviously, different hazardous areas may have different causes. Therefore, the identification and segmentation of hazardous areas are two essential data preprocessing procedures for reliability enhancement.
\begin{figure}[!t]
\centering
\includegraphics[height=1.80in,width=3.25in]{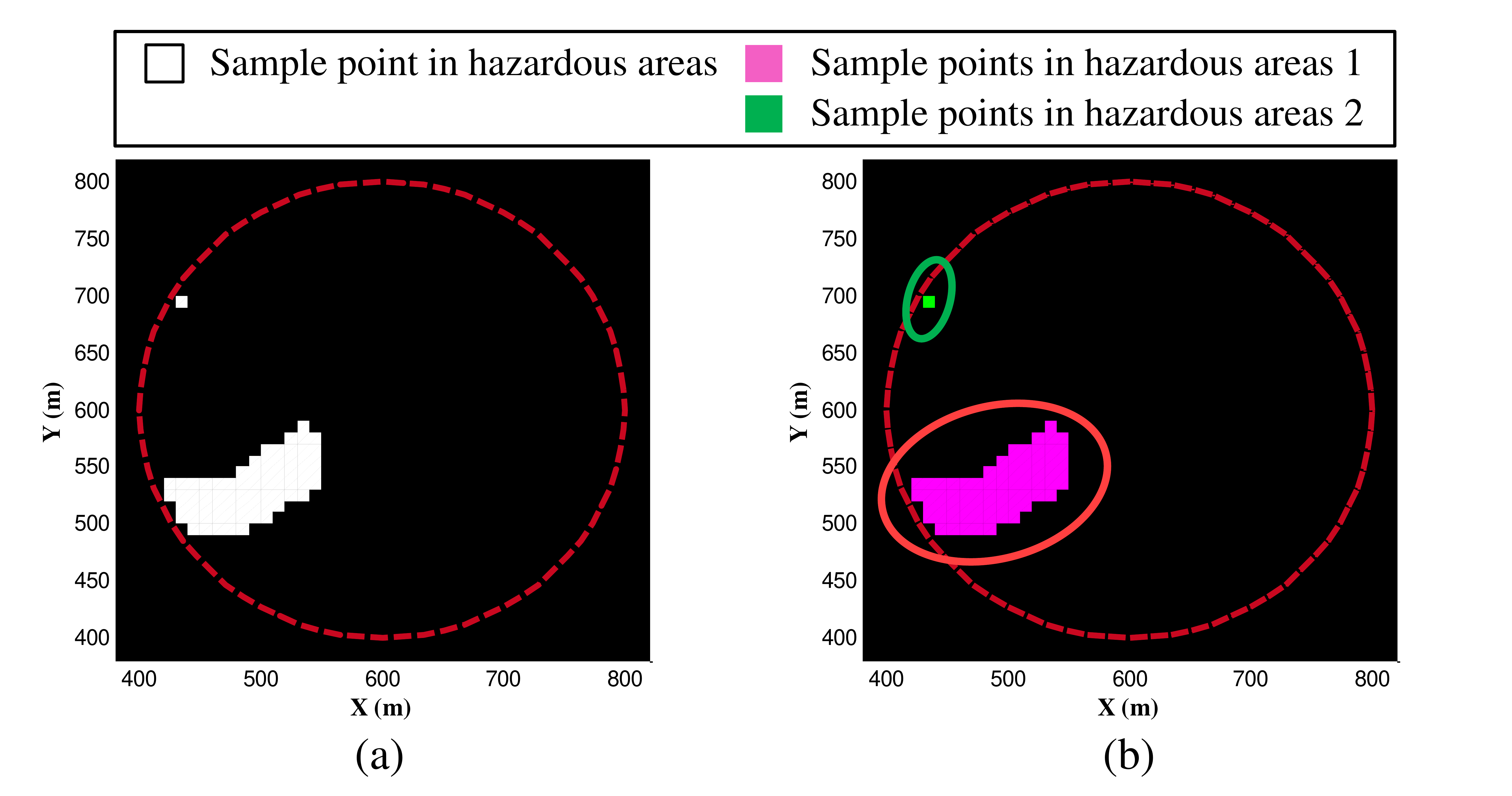}
\caption{Hazardous areas (a) identification and (b) segmentation.}
\label{fig_15)}
\end{figure}

In this article, the identification of hazardous areas is realized by comparing the reliability prediction result (${\eta ^{ * m}}$) of each sample point $m$ with a predetermined threshold ${\eta _T}$. The decision rule can be expressed as
\begin{equation}\tag{64}
m \in \left\{ {\begin{array}{*{20}{c}}
\begin{aligned}
&{{\cal M}_H},&&{\rm{if}}\ {\eta ^{ * m}} > {\eta _T},\\
&{{\cal M}_0},&&{\rm{otherwise}},
\end{aligned}
\end{array}} \right.
\end{equation}
where ${{\cal M}_H}$ and ${{\cal M}_0}$ are sets composed of sample points inside and outside the hazardous areas, respectively. It is worth noting that the threshold ${\eta _T}$ is usually set to be slightly smaller than ${\eta _{REQ}}$, that is, ${\eta _T} < {\eta _{REQ}}$. The reason for this is to provide a margin against the reliability degradation outside the hazardous areas caused by the adjustment of SPs' locations. Fig. 12(a) shows the result of hazardous area identification.

After finishing the identification process, we consider how to determine which area a sample point belongs to, that is, the segmentation of hazardous area. In this article, the well-known 8-connected neighborhood approach is used to determine the boundary of each hazardous area and segment sample points. The segmentation results are shown in Fig. 12(b).

\subsection{Proposed Voting-based Cause Analysis Method for Reliability Enhancement}
Intuitively speaking, the major cause of hazardous areas in the test scenario is that the mountain surrounded by yellow circle in Fig. 11(b) reduces the visibility of several SPs and increases the probability of NLoS propagation. However, this kind of intuitive analysis is not sufficient to provide guidance for the adjustment of SPs' locations and is not suitable for complex terrain environments. In order to achieve the goal of eliminating hazardous areas in the reliability prediction results, it is extremely important to find out the specific cause of each hazardous area. Specifically, we need to know which SPs' low visibility or high failure rates lead to the existence of a certain hazardous area.
\begin{algorithm}[!t]
\caption{Voting-based Method for the Cause Analysis of Hazardous Area $i$}
\label{alg:Voting Method}
\begin{algorithmic}[1]
\renewcommand{\algorithmicrequire}{ \textbf{Input:}}
\REQUIRE
The set ${{\cal M}_i}$ composed of sample points in i-th hazardous area, reliability prediction results (${\eta ^{ * {m_i},x}}$ and ${\eta ^{ * {m_i},y}}$) corresponding to each sample point ${m_i}$ (${m_i} \!\in\! {{\cal M}_i}$).
\renewcommand{\algorithmicrequire}{ \textbf{Initialization:}}
\REQUIRE ~\\
\STATE Initialize voting vectors: ${\bf{v}}_{{H_i},U}^{*x} \!\leftarrow\! {{\bf{0}}_{1 \times K}}$, ${\bf{v}}_{{H_i},\left. F \right|O}^{*x} \!\leftarrow\! {{\bf{0}}_{1 \times K}}$, ${\bf{v}}_{{H_i},U}^{*y} \!\leftarrow\! {{\bf{0}}_{1 \times K}}$, ${\bf{v}}_{{H_i},\left. F \right|O}^{*y} \!\leftarrow\! {{\bf{0}}_{1 \times K}}$;\\
\renewcommand{\algorithmicrequire}{ \textbf{Voting Process:}}
\REQUIRE ~\\
\FOR{${m_i}=1$ to ${M_i}$}
\STATE Calculate the prior probabilities $P_{k,B}^{m_i}$ and $P_{\left.\! {k,F} \!\right|O}^{m_i}$;\\
\STATE Find the largest number of unavailable SPs in the remaining observation events: ${\bar U^{ * {m_i}}} = \mathop {\max }\limits_{g \in {{\cal G}^{m_i}}} \left\{ {{{\bar U}_g}} \right\}$;\\
\STATE Find the largest number of failure SPs in the remaining failure events: ${\bar F^{ * {m_i}}} = \mathop {\max }\limits_{g \in {{\cal G}^{m_i}}} \left\{ {\mathop {\max }\limits_{q \in \bar {\cal Q}_g^{m_i}} \left\{ {\bar F_q^g} \right\}} \right\}$;\\
\IF{${\eta ^{*{m_i},x}} \!>\! {\eta _T}$}
\FOR{${k}=1$ to ${K}$}
\IF{$P_{k,B}^{m_i}$ is the largest ${\bar U^{ * {m_i}}}$ among $K$ SPs}
\STATE ${\bf{v}}_{{H_i},U}^{*x}\left[ k \right] \!\leftarrow\! {\bf{v}}_{{H_i},U}^{*x}\left[ k \right] \!+\! \frac{{{{\left( {{\eta ^{*{m_i},x}} \!-\! {\eta _T}} \right)}^2}}}{{\sum\limits_{{m_i} \in {{\cal M}_i}} {{{\left( {{\eta ^{*{m_i},x}} \!-\! {\eta _T}} \right)}^2}} }}$;\\
\ENDIF
\IF{$P_{\left.\! {k,F} \!\right|O}^{m_i}$ is the largest ${\bar F^{ * {m_i}}}$ among $K$ SPs}
\STATE ${\bf{v}}_{{H_i},\left. F \right|O}^{*x}\left[ k \right] \!\leftarrow\! {\bf{v}}_{{H_i},\left. F \right|O}^{*x}\left[ k \right] \!+\! \frac{{{{\left( {{\eta ^{*{m_i},x}} \!-\! {\eta _T}} \right)}^2}}}{{\sum\limits_{{m_i} \in {{\cal M}_i}} {{{\left( {{\eta ^{*{m_i},x}} \!-\! {\eta _T}} \right)}^2}} }}$;\\
\ENDIF
\ENDFOR
\ENDIF
\STATE Repeat steps 6-15 in the y-direction;\\
\ENDFOR
\STATE ${\bf{v}}_{{H_i},U}^{*x} \!\leftarrow\! \left\langle{\bf{v}}_{{H_i},U}^{*x}\right\rangle$, ${\bf{v}}_{{H_i},\left. F \right|O}^{*x} \!\leftarrow\! \left\langle{\bf{v}}_{{H_i},\left. F \right|O}^{*x}\right\rangle$, ${\bf{v}}_{{H_i},U}^{*y} \!\leftarrow\! \left\langle{\bf{v}}_{{H_i},U}^{*y}\right\rangle$, ${\bf{v}}_{{H_i},\left. F \right|O}^{*y} \!\leftarrow\! \left\langle{\bf{v}}_{{H_i},\left. F \right|O}^{*y}\right\rangle$;\\
\renewcommand{\algorithmicrequire}{ \textbf{Output:}}
\REQUIRE
${\bf{v}}_{{H_i},U}^{*x}$, ${\bf{v}}_{{H_i},\left. F \right|O}^{*x}$, ${\bf{v}}_{{H_i},\left. F \right|O}^{*y}$ and ${\bf{v}}_{{H_i},\left. F \right|O}^{*y}$.
\end{algorithmic}
\end{algorithm}

It is worth noting that even the cause analysis of a single hazardous area is not easy to implement. For example, hazardous area 1 in Fig. 11(c) covers a large area and contains a number of sample points. The reasons for the unsatisfactory reliability of these sample points are very unlikely to be the same. However, it is obviously impossible to adjust SPs' locations according to the requirements of each sample point. To address this problem, we propose a voting-based cause analysis method, which can fuse the causes of all sample points inside a hazardous area into a final cause analysis result corresponding to the whole area. The proposed method is summarized in \emph{Algorithm 4}. The key idea of this method is to let each sample point vote for the SPs that have the greatest impacts on its reliability, and the SPs with weighted votes more than a threshold value (0.5) will be regarded as the cause of the hazardous area. As can be seen from \emph{Algorithm 4}, there are four types of voting vectors in the proposed method, namely, ${\bf{v}}_{{H_i},U}^{ * x}$, ${\bf{v}}_{{H_i},U}^{ * y}$, ${\bf{v}}_{{H_i},\left. F \right|O}^{ * x}$ and ${\bf{v}}_{{H_i},\left. F \right|O}^{ * y}$. ${\bf{v}}_{{H_i},U}^{ * x}\left[ k \right] = 1$ means that the low visibility of the k-th SP is one of the reasons for the i-th hazardous area's unsatisfactory reliability in x-direction. ${\bf{v}}_{{H_i},\left. F \right|O}^{ * x}\left[ k \right] = 1$ indicates that the high conditional failure rate of the k-th SP leads to the poor reliability of the i-th hazardous area in x-direction. ${\bf{v}}_{{H_i},U}^{ * y}$ and ${\bf{v}}_{{H_i},\left. F \right|O}^{ * y}$ have similar meanings to ${\bf{v}}_{{H_i},U}^{ * x}$ and ${\bf{v}}_{{H_i},\left. F \right|O}^{ * x}$, respectively. Table I shows the voting results corresponding to the test scenario. Note that the voting vectors in Table I have been binarized in the last step of \emph{Algorithm 4}. It can be concluded that the low visibility of SPs 3, 4 and 5 is the major cause of hazardous area 1. The low visibility of SPs 6, 7 and the high conditional failure rate of SP 5 lead to the poor reliability of hazardous area 2.
\begin{table}[!t]
\renewcommand\arraystretch{1.0}
\caption{Voting Results in the Cause Analysis of Hazardous Areas}
\label{Vote_Table}
\centering
\begin{tabular}{ccccccccc}
\toprule
& SP1 & SP2 & SP3 & SP4 & SP5 & SP6 & SP7 & SP8 \\
\midrule
${\bf{v}}_{{H_1},U}^{ * x}$ & 0 & 0 & 0 & 1 & 1 & 0 & 0 & 0 \\
\midrule
${\bf{v}}_{{H_1},\left. F \right|O}^{ * x}$ & 0 & 0 & 0 & 1 & 1 & 0 & 0 & 0 \\
\midrule
${\bf{v}}_{{H_1},U}^{ * y}$ & 0 & 0 & 1 & 1 & 0 & 0 & 0 & 0 \\
\midrule
${\bf{v}}_{{H_1},\left. F \right|O}^{ * x}$ & 0 & 0 & 1 & 1 & 1 & 0 & 0 & 0 \\
\midrule
${\bf{v}}_{{H_2},U}^{ * x}$ & 0 & 0 & 0 & 0 & 0 & 0 & 0 & 0 \\
\midrule
${\bf{v}}_{{H_2},\left. F \right|O}^{ * x}$ & 0 & 0 & 0 & 0 & 0 & 0 & 0 & 0 \\
\midrule
${\bf{v}}_{{H_2},U}^{ * y}$ & 0 & 0 & 0 & 0 & 0 & 1 & 1 & 0 \\
\midrule
${\bf{v}}_{{H_2},\left. F \right|O}^{ * x}$ & 0 & 0 & 0 & 0 & 1 & 1 & 0 & 0 \\
\bottomrule
\end{tabular}
\end{table}
\begin{figure}[!t]
\centering
\includegraphics[height=3.15in,width=3.45in]{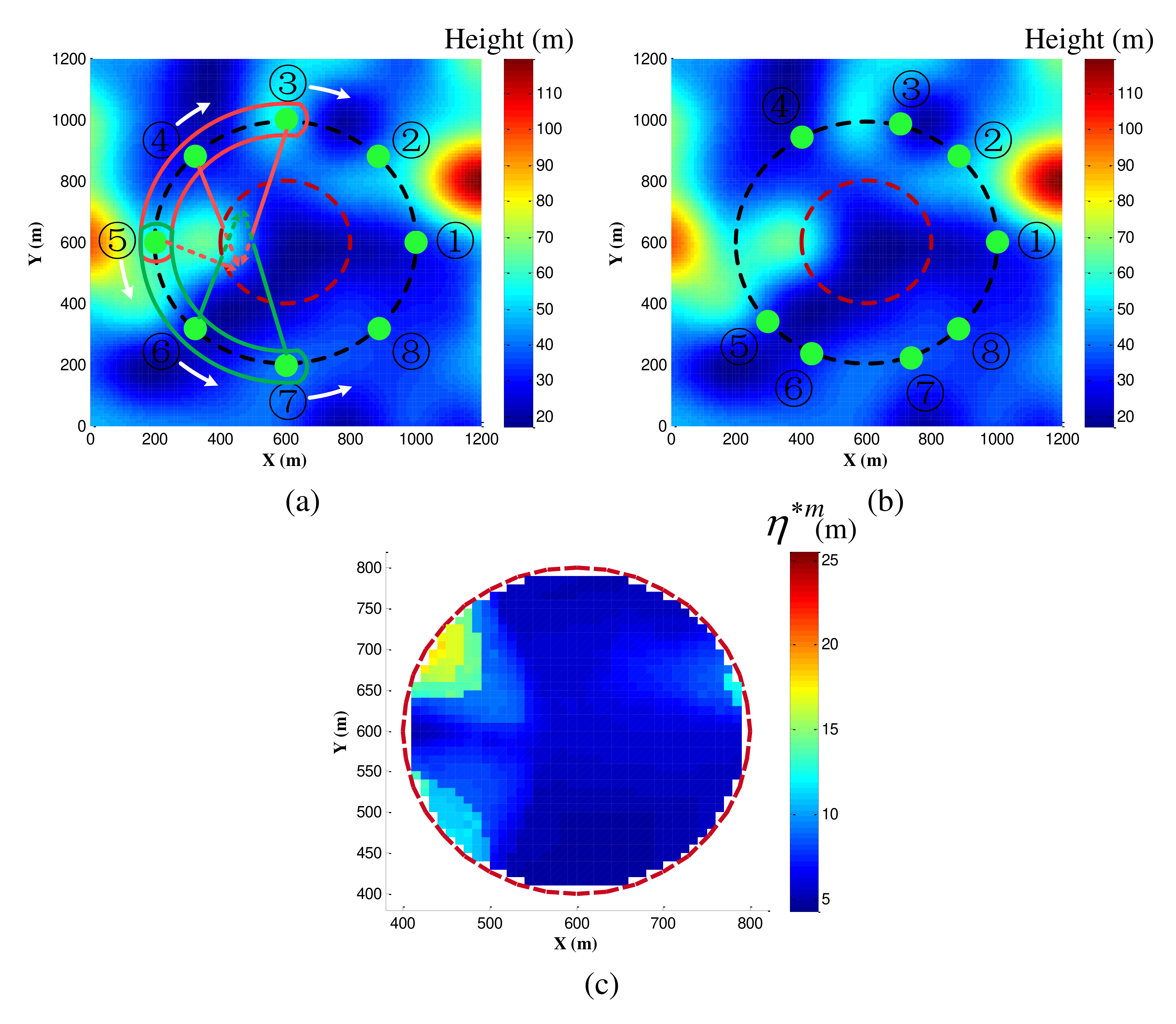}
\caption{Example of reliability enhancement: (a) Guidance for SP location adjustment, (b) adjusted SPs, and (c) the enhancement results (${\eta ^{ * m}}$).}
\label{fig_16)}
\end{figure}

The cause analysis results obtained by the proposed method provide helpful guidance for the adjustment of SPs locations. For hazardous area 1, its reliability is limited by the visibility of SPs 3 and 4. Thus, in order to improve the visibility of these two SPs in hazardous area 1, we rotate them clockwise around the center of the location uncertain area by appropriate angles, as shown in Fig. 13(a). Similarly, we rotate SPs 6 and 7 counterclockwise to improve their visibility in hazardous area 2. Since SP 5 affects the reliability of both areas, the adjustment of its location is the most complicated. To make matter worse, the two hazardous areas have conflicting requirements for the adjustment of SP 5. For example, rotating SP 5 clockwise could reduce its conditional failure rate in hazardous area 1, but at the same time further degrade its visibility in hazardous area 2. It is noteworthy that the cause analysis results shown in Table I can also help us solve this problem. It can be seen from Table I, SP 5 has good visibility but high conditional failure rate in hazardous area 2. For this type of SPs, in addition to reducing its conditional failure rate, greatly reducing its visibility to make it almost unavailable is also a potential way to reduce its impact on reliability. Therefore, we rotate SP 5 counterclockwise to improve and reduce its visibility in hazardous area 1 and 2, respectively. The adjusted SP locations and the enhanced reliability prediction results are shown in Fig. 13(b) and (c), respectively.

As can be seen from Fig. 13(c), the enhanced reliability meets the mission's requirement (${\eta ^{ * m}} \!<\! {\eta _{REQ}} \!=\! 20{\rm{m}}$), which demonstrates the effectiveness of the proposed voting-based method. Currently, the proposed method only provides guidance for SP adjustment, while the exact locations of adjusted SPs are manually selected according to the guidance. In our future work, we will try to develop a novel reliability enhancement method that could perform SP adjustment automatically.

\section{Numerical Results}
\begin{figure}[!t]
\centering
\includegraphics[height=1.85in,width=3.45in]{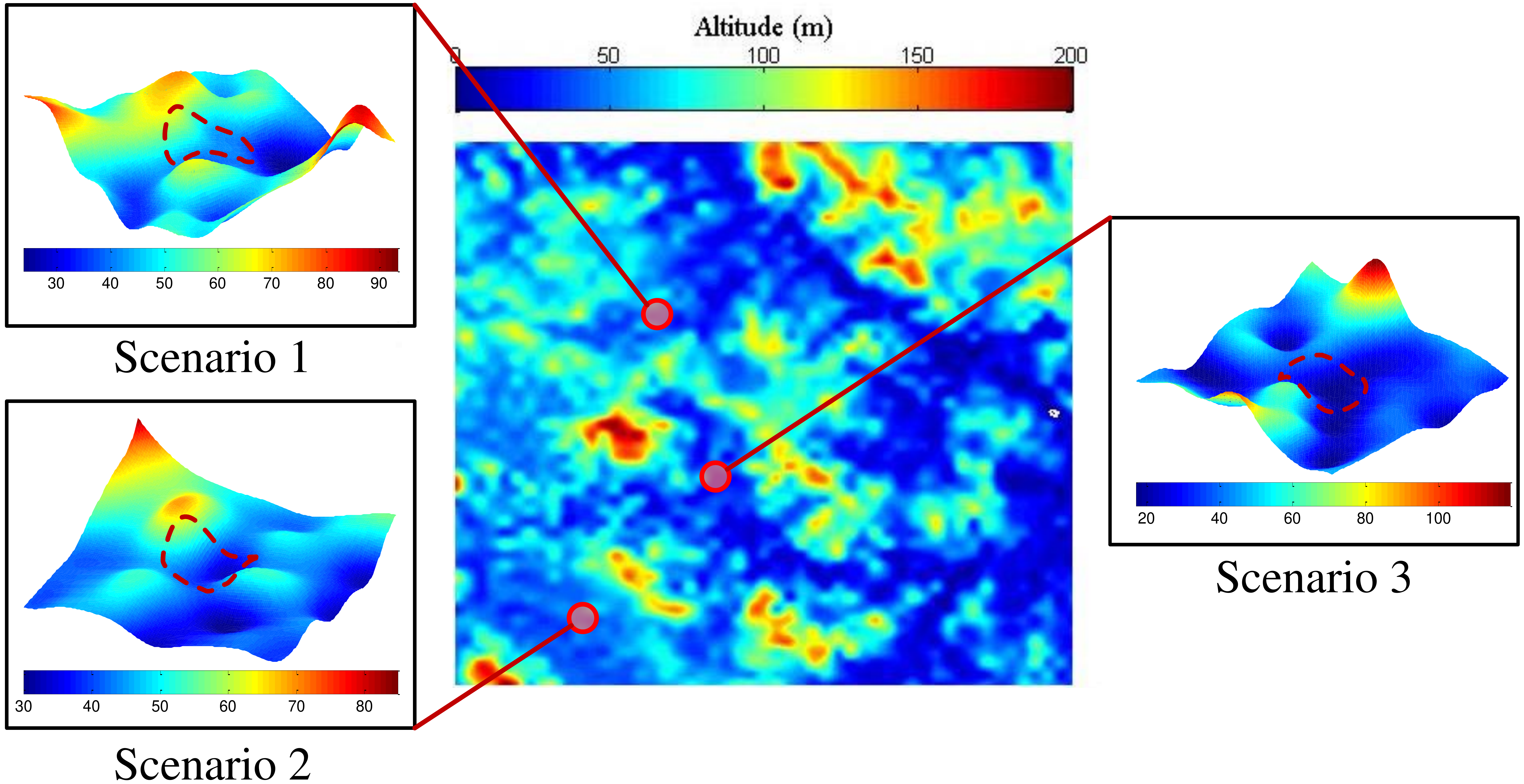}
\caption{Test scenarios.}
\label{fig_13(A))}
\end{figure}
In this section, a series of simulation experiments are conducted and the corresponding numerical results are provided to verify the feasibility of the proposed system, as well as the validity and performance of the proposed methods. As shown in Fig. 14, we selected three test scenarios from a typical mountainous environment. Among them, scenario 3 has been used in the previous section, and the remaining two scenarios will be used for experiments in this section. First, we test the service reliability of the proposed UAV-enabled positioning system in scenario 1 with the proposed reliability prediction method, and analyze the key factors affecting reliability. Then, the effectiveness of the proposed voting-based method for reliability enhancement is demonstrated via an experiment conducted in scenario 2. Table II summarizes the key simulation parameters used in this section.
\begin{table}[!t]
\renewcommand{\arraystretch}{1.5}
\newcommand{\tabincell}[2]{\begin{tabular}{@{}#1@{}}#2\end{tabular}}
\caption{Simulation Parameters}
\label{table_comp}
\centering
\begin{tabular}{|l|l|}
\hline
Parameter&Value \\
\hline
Main frequency (${f_c}$)& 1.5 GHz \\
\hline
Path loss exponent (PLE) in NLoS (${\alpha _N}$)& 3.4 \\
\hline
Standard deviation of shadowing (${\sigma _N}$)& 1.4 dB \\
\hline
Transmission power of user device (${P_{t,U}}$)& 20 dBm \\
\hline
Noise power (${P_{{n_0}}}$)& -104 dBm \\
\hline
Standard deviation of terrain uncertainty (${\sigma _h}$)& 1 m \\
\hline
Radius of location uncertainty area (${R_{Un}}$)& 200 m \\
\hline
Number of sample points ($M$)& 1257 (interval 10m) \\
\hline
Number of service points (SPs) ($K$)& 8 \\
\hline
UAV altitude (${h_B}$)& 100 m \\
\hline
\tabincell{l}{Minimum horizontal distance between SP and\\ the center of location uncertainty area (${d_{\min }}$)}& 400 m \\
\hline
Response delay of TWR (${\tau _D}$)& 5 ms \\
\hline
Crystal tolerance of user's oscillator (${O_U}$)& 10 ppm \\
\hline
Prior probability of internal fault (${P_{IF}}$)& ${10^{ - 6}}$ \\
\hline
Tolerable limit of position error (${\eta _{REQ}}$)& 20 m \\
\hline
Tolerable FA rate ($P_{FA}^{REQ}$)& ${10^{ - 4}}$ \\
\hline
Tolerable MD rate ($P_{MD}^{REQ}$)& ${10^{ - 6}}$ \\
\hline
Threshold for hazardous areas (${\eta _T}$)& 18 m \\
\hline
\end{tabular}
\end{table}

\subsection{Reliability Test of the Proposed System}
\begin{figure}[!t]
\centering
\includegraphics[height=1.65in,width=3.45in]{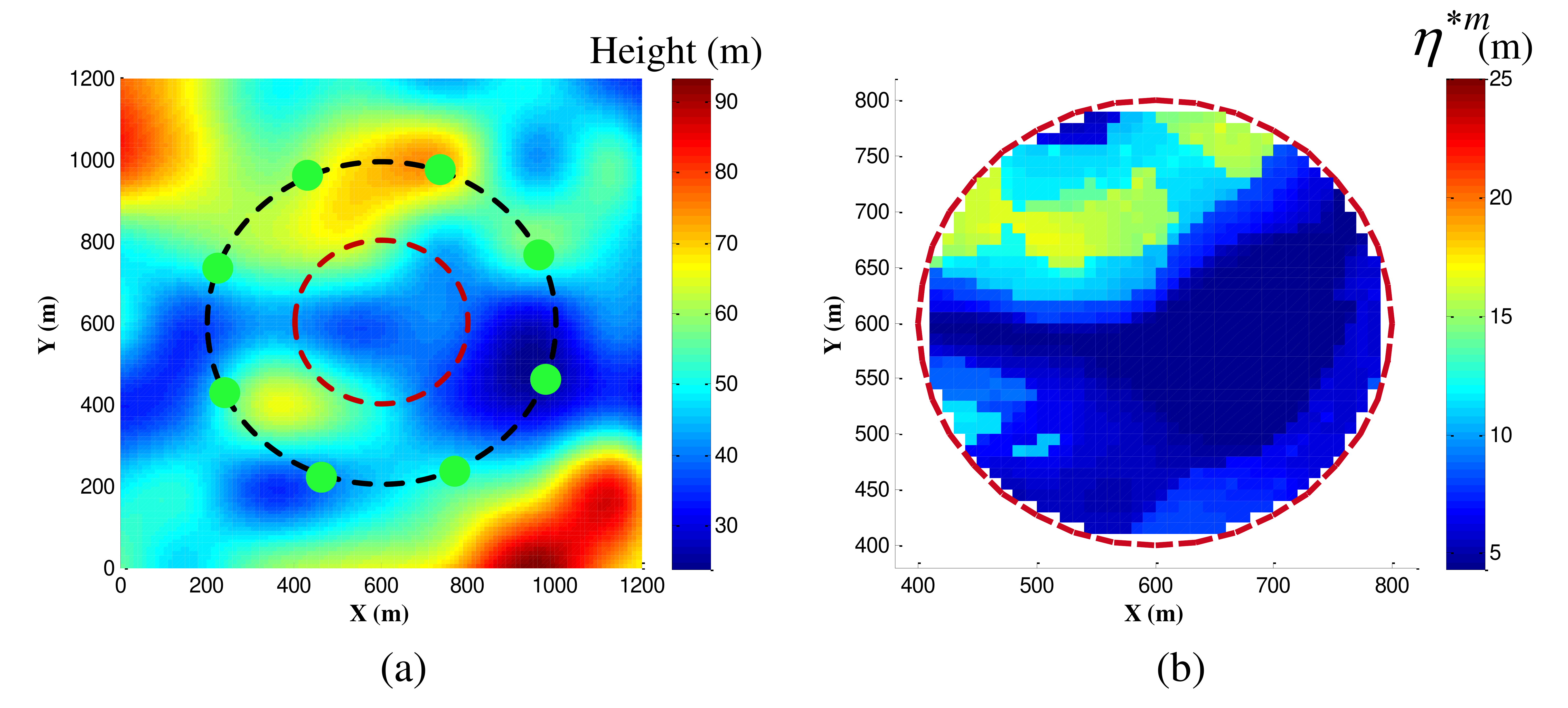}
\caption{Reliability prediction in scenario 1: (a) Top view of test scenario 1 and (b) the corresponding prediction results (${\eta ^{ * m}}$).}
\label{fig_17)}
\end{figure}
Utilizing the proposed reliability prediction method described in Section III, we test the service reliability of the proposed system in the test scenario shown in Fig. 15(a), where a user is located in a valley surrounded by mountains. This scenario is considered to be very challenging for conventional GNSS systems or terrestrial cellular-based positioning technologies, because their signals are very likely to be blocked by terrain. The reliability prediction results are shown in Fig. 15(b). It can be seen that in such a challenging environment, the predicted minimum detectable position error of the proposed system is 16.8m, which meets the mission's requirement for service reliability (${\eta _{REQ}} = 20{\rm{m}}$). In fact, such performance is sufficient to support many life-critical missions like field rescue and disaster management. Therefore, it can be concluded that the proposed system has the potential to provide highly reliable positioning service for ground users in mountainous environments.

We further analyze the key factors affecting the reliability of the proposed system. Obviously, the altitude of the UAV platform affects the reliability. The red line in Fig. 16 shows the variation of the predicted reliability with the UAV altitude. It can be seen that the reliability improves as the altitude increases. This phenomenon is quite easy to understand since the increase in altitude leads to better visibility as well as a smaller NLoS probability, which ultimately results in better reliability. However, it should be noted that in practical applications, UAV altitude is restricted by air traffic control and safety considerations, which means that we should not always try to improve the reliability by increasing the altitude.

Moreover, we also study the influence of SPs' locations on reliability. In the experiment, we change the locations of SPs by rotating all of them counterclockwise at the same time. The bar chart in Fig. 16 shows the reliability prediction results obtained at different rotation angles. It can be seen that with the increase of rotation angle, the predicted minimum detectable error first increases and then decreases, ranging from 16.8m to 27.1m. The reliability corresponding to the best rotation angle (${5^ \circ }$) has a 38$\%$ improvement compared with that corresponding to the worst rotation angle (${20^ \circ }$). The above phenomenon indicates that the selection of SPs' locations is also one of the main factors that affect the proposed system's reliability. In addition, optimizing the locations of SPs could improve the service reliability without changing the UAV altitude, which is an important basis for reliability enhancement.
\begin{figure}[!t]
\centering
\includegraphics[height=2.10in,width=3.35in]{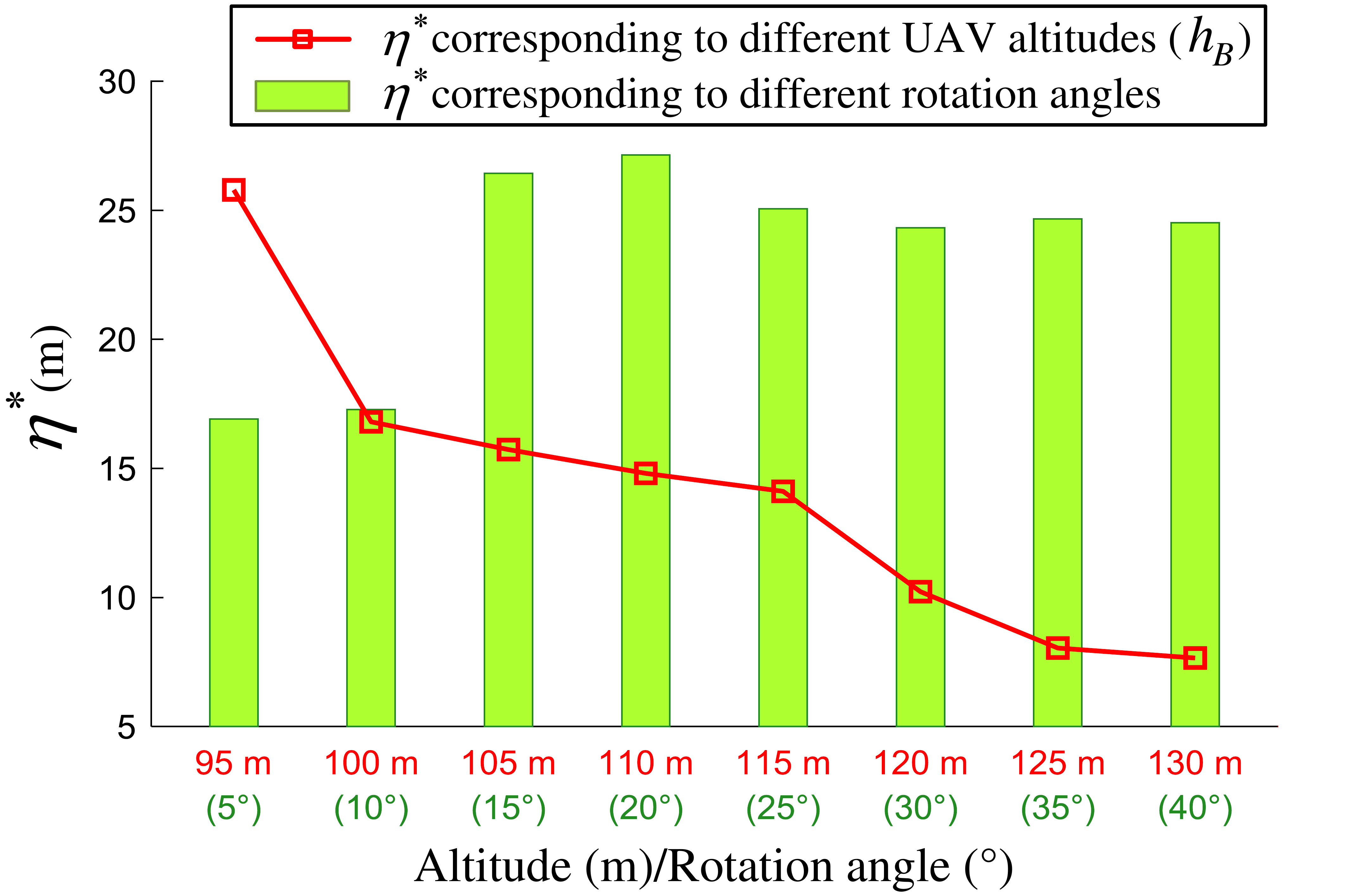}
\caption{Analysis of key factors affecting reliability.}
\label{fig_18)}
\end{figure}

\subsection{Demonstration of the Proposed Enhancement Method}
The proposed voting-based cause analysis method for reliability enhancement has been tested in Section IV. In this subsection, we test it again in scenario 2 to demonstrate its effectiveness in different scenarios. The test scenario and its corresponding reliability prediction results are shown in Fig. 17(a) and (b), respectively. It can be seen that there is a hazardous area in Fig. 17(b). Then, we use the proposed method to analyze the cause of this hazardous area, and the analysis results are shown in Table III. The analysis results indicate that the low visibility of SPs 2, 3 and the high conditional failure rate of SP 1 lead to the unsatisfactory reliability of the hazardous area.

Based on the cause analysis results obtained by the proposed method, we adjust the locations of SPs as follows. In order to improve the visibility of SPs 2 and 3, we rotate them counterclockwise by appropriate angles so that their signals are less likely to be blocked by mountains. Then, we rotate SP 1 clockwise to reduce its NLoS probability. The adjusted SP locations and the corresponding enhanced reliability are shown in Fig. 17(c) and (d), respectively. It can be seen that the enhanced reliability meets the mission's requirement, which demonstrates that the proposed method is effective for reliability enhancement in mountainous environments.
\begin{figure}[!t]
\centering
\includegraphics[height=3.15in,width=3.45in]{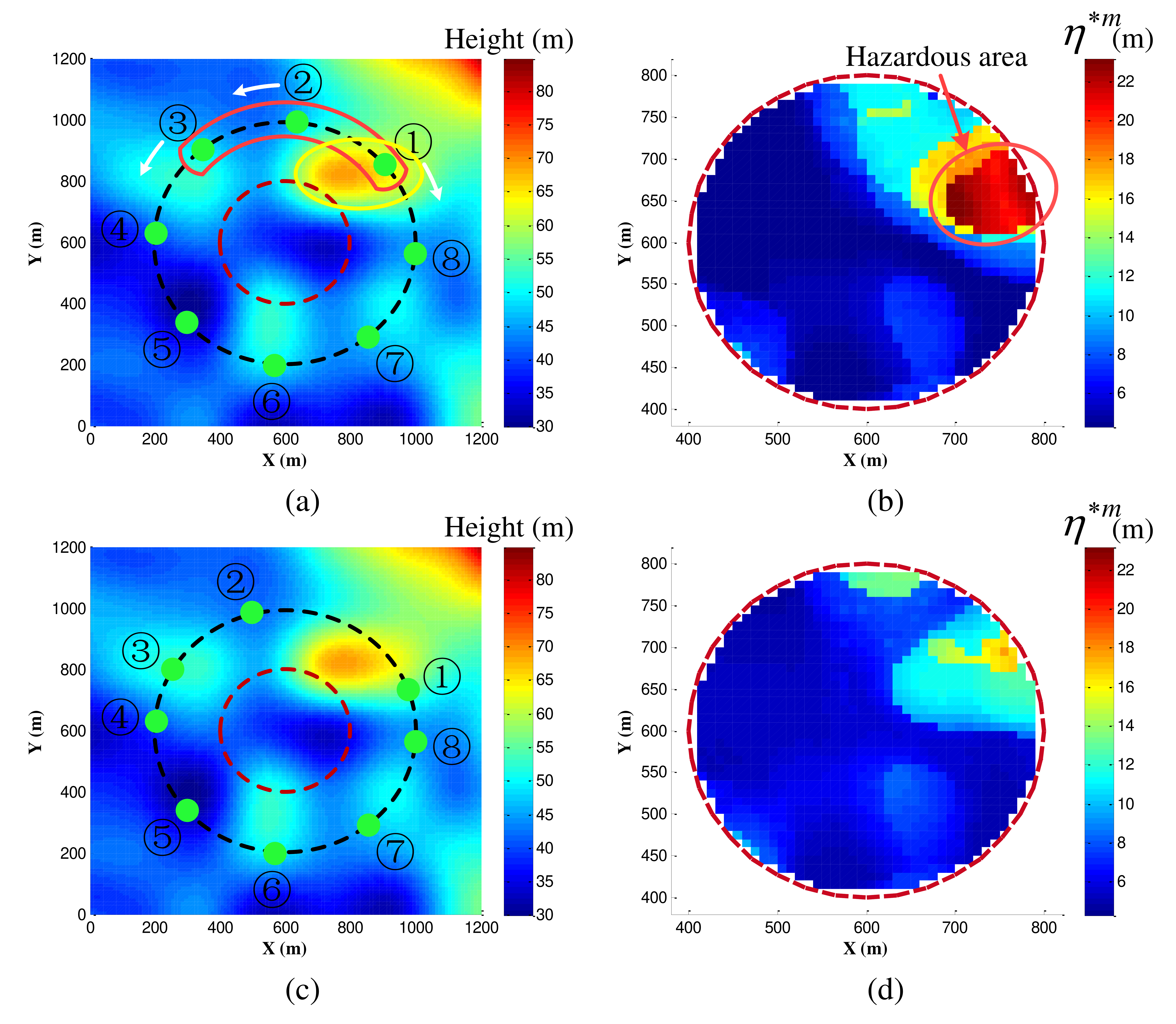}
\caption{Reliability prediction and enhancement in scenario 2: (a) Top view of test scenario 2 and (b) the prediction results (${\eta ^{ * m}}$), (c) adjusted SPs and (d) the enhancement results (${\eta ^{ * m}}$).}
\label{fig_19)}
\end{figure}
\begin{table}[!t]
\renewcommand\arraystretch{1.0}
\caption{Cause Analysis Results in Test Scenario 2}
\label{Table II}
\centering
\begin{tabular}{ccccccccc}
\toprule
& SP1 & SP2 & SP3 & SP4 & SP5 & SP6 & SP7 & SP8 \\
\midrule
${\bf{v}}_{{H_1},U}^{ * x}$ & 0 & 1 & 1 & 0 & 0 & 0 & 0 & 0 \\
\midrule
${\bf{v}}_{{H_1},\left. F \right|O}^{ * x}$ & 1 & 1 & 0 & 0 & 0 & 0 & 0 & 0 \\
\midrule
${\bf{v}}_{{H_1},U}^{ * y}$ & 0 & 1 & 1 & 0 & 0 & 0 & 0 & 0 \\
\midrule
${\bf{v}}_{{H_1},\left. F \right|O}^{ * x}$ & 1 & 1 & 0 & 0 & 0 & 0 & 0 & 0 \\
\bottomrule
\end{tabular}
\end{table}

\section{Conclusion}
This article presents the progress made in the first phase of our research project on reliable UAV-enabled positioning, which aims to develop a novel system that uses a low-altitude UAV platform to provide highly reliable positioning services in mountainous environments. In this phase, we solved the reliability prediction problem in our UAV-enabled positioning system, and conducted a preliminary study on reliability enhancement. We first designed the structure and operation scheme of the system, and selected the appropriate method to support positioning service. Then, the major causes of service failures were theoretically analyzed, and a failure model was established to describe their prior probabilities and impacts on positioning performance. Based on the established failure model, we propose a reliability prediction method to evaluate the reliability of the positioning service before the UAV takes off. Moreover, for those situations where the predicted reliability fails to meet the requirements, we also proposed a voting-based method to analysis the causes of unsatisfactory reliability and provide guidance for reliability enhancement. Numerical results demonstrated the tremendous potential of our system in reliable positioning, as well as the effectiveness of the proposed methods. We hope that this article would bring inspiration for the application of UAVs in future networks and lead to a new practical solution for reliable positioning service.

\bibliographystyle{IEEEtran}
\bibliography{mybib}

\begin{thebibliography}{10}
\providecommand{\url}[1]{#1}
\csname url@samestyle\endcsname
\providecommand{\newblock}{\relax}
\providecommand{\bibinfo}[2]{#2}
\providecommand{\BIBentrySTDinterwordspacing}{\spaceskip=0pt\relax}
\providecommand{\BIBentryALTinterwordstretchfactor}{4}
\providecommand{\BIBentryALTinterwordspacing}{\spaceskip=\fontdimen2\font plus
\BIBentryALTinterwordstretchfactor\fontdimen3\font minus
  \fontdimen4\font\relax}
\providecommand{\BIBforeignlanguage}[2]{{%
\expandafter\ifx\csname l@#1\endcsname\relax
\typeout{** WARNING: IEEEtran.bst: No hyphenation pattern has been}%
\typeout{** loaded for the language `#1'. Using the pattern for}%
\typeout{** the default language instead.}%
\else
\language=\csname l@#1\endcsname
\fi
#2}}
\providecommand{\BIBdecl}{\relax}
\BIBdecl

\bibitem{LBS_Intr_1}
A.~K{\"u}pper, \emph{Location-based services: fundamentals and
  operation}.\hskip 1em plus 0.5em minus 0.4em\relax John Wiley $\&$ Sons,
  2005.

\bibitem{TR_38.855}
{3GPP TR 38.855}, ``Study on {NR} positioning support,'' Rel. 16, 2019.

\bibitem{IoT_LBS}
A.~{Hoglund}, X.~{Lin}, O.~{Liberg}, A.~{Behravan}, E.~A. {Yavuz}, M.~{Van Der
  Zee}, Y.~{Sui}, T.~{Tirronen}, A.~{Ratilainen}, and D.~{Eriksson}, ``Overview
  of {3GPP} release 14 enhanced {NB-IoT},'' \emph{IEEE Network}, vol.~31,
  no.~6, pp. 16--22, 2017.

\bibitem{5G_Mount}
J.~{Lee}, E.~{Tejedor}, K.~{Ranta-aho}, H.~{Wang}, K.~{Lee}, E.~{Semaan},
  E.~{Mohyeldin}, J.~{Song}, C.~{Bergljung}, and S.~{Jung}, ``Spectrum for
  {5G}: Global status, challenges, and enabling technologies,'' \emph{IEEE
  Communications Magazine}, vol.~56, no.~3, pp. 12--18, 2018.

\bibitem{IoT_Mount}
T.~O. {Olasupo}, ``Wireless communication modeling for the deployment of tiny
  {IoT} devices in rocky and mountainous environments,'' \emph{IEEE Sensors
  Letters}, vol.~3, no.~7, pp. 1--4, 2019.

\bibitem{Mount_Terr}
M.~{Lindong}, S.~{Bo}, X.~{Peng}, L.~{Zhenjie}, J.~{Chenxing}, D.~{Ruina}, and
  D.~{Qindan}, ``A strategy to climb vertical cliffs for the quadruped robot
  imitating goat,'' in \emph{2018 3rd International Conference on Control,
  Robotics and Cybernetics (CRC)}, 2018, pp. 37--41.

\bibitem{Pos_Disa}
G.~{Han}, X.~{Yang}, L.~{Liu}, W.~{Zhang}, and M.~{Guizani}, ``A disaster
  management-oriented path planning for mobile anchor node-based localization
  in wireless sensor networks,'' \emph{IEEE Transactions on Emerging Topics in
  Computing}, vol.~8, no.~1, pp. 115--125, 2020.

\bibitem{Phone_GNSS_1}
X.~Zhang, X.~Tao, F.~Zhu, X.~Shi, and F.~Wang, ``Quality assessment of {GNSS}
  observations from an android {N} smartphone and positioning performance
  analysis using time-differenced filtering approach,'' \emph{GPS Solutions},
  vol.~22, no.~3, p.~70, 2018.

\bibitem{Phone_GNSS_2}
P.~Dabove and V.~{Di Pietra}, ``Towards high accuracy {GNSS} real-time
  positioning with smartphones,'' \emph{Advances in Space Research}, vol.~63,
  no.~1, pp. 94 -- 102, 2019.

\bibitem{OTDoA_Rel9}
S.~Fischer, ``Observed time difference of arrival ({OTDOA}) positioning in
  {3GPP LTE},'' \emph{Qualcomm White Pap}, 2014.

\bibitem{LTE_Urban}
K.~Shamaei and Z.~M. Kassas, ``{LTE} receiver design and multipath analysis for
  navigation in urban environments,'' \emph{Navigation}, vol.~65, no.~4, pp.
  655--675, 2018.

\bibitem{LTE_Indoor}
M.~{Driusso}, C.~{Marshall}, M.~{Sabathy}, F.~{Knutti}, H.~{Mathis}, and
  F.~{Babich}, ``Indoor positioning using {LTE} signals,'' in \emph{2016
  International Conference on Indoor Positioning and Indoor Navigation (IPIN)},
  2016, pp. 1--8.

\bibitem{GNSS_Block}
L.~Peraza, M.~Semmling, C.~Falck, O.~Pavlova, S.~Gerland, and J.~Wickert,
  ``Analysis of grazing {GNSS} reflections observed at the zeppelin mountain
  station, spitsbergen,'' \emph{Radio Science}, vol.~52, no.~11, pp.
  1352--1362, 2017.

\bibitem{GNSS_Num}
W.~{Jiang}, D.~{Liu}, B.~{Cai}, C.~{Rizos}, J.~{Wang}, and W.~{Shangguan}, ``A
  fault-tolerant tightly coupled {GNSS/INS/OVS} integration vehicle navigation
  system based on an {FDP} algorithm,'' \emph{IEEE Transactions on Vehicular
  Technology}, vol.~68, no.~7, pp. 6365--6378, 2019.

\bibitem{GNSS_NLoS}
R.~{Rahdar}, J.~T. {Stracener}, and E.~V. {Olinick}, ``A systems engineering
  approach to improving the accuracy of mobile station location estimation,''
  \emph{IEEE Systems Journal}, vol.~8, no.~1, pp. 14--22, 2014.

\bibitem{GNSS_Geo}
D.~{Lu}, S.~{Jiang}, B.~{Cai}, W.~{Shangguan}, X.~{Liu}, and J.~{Luan},
  ``Quantitative analysis of {GNSS} performance under railway obstruction
  environment,'' in \emph{2018 IEEE/ION Position, Location and Navigation
  Symposium (PLANS)}, 2018, pp. 1074--1080.

\bibitem{LTE_Chal}
Q.~{Liu}, R.~{Liu}, Z.~{Wang}, and Y.~{Zhang}, ``Simulation and analysis of
  device positioning in {5G} ultra-dense network,'' in \emph{2019 15th
  International Wireless Communications Mobile Computing Conference (IWCMC)},
  2019, pp. 1529--1533.

\bibitem{LTE_NLoS}
K.~Shamaei and Z.~M. Kassas, ``{LTE} multipath mitigation for positioning in
  urban environments,'' \emph{Navigation: Journal of the Institute of
  Navigation}, 2017.

\bibitem{UAV_RTK}
Z.~Liu, Z.~Li, B.~Liu, X.~Fu, I.~Raptis, and K.~Ren, ``Rise of mini-drones:
  Applications and issues,'' in \emph{Proceedings of the 2015 Workshop on
  Privacy-Aware Mobile Computing}, ser. PAMCO '15.\hskip 1em plus 0.5em minus
  0.4em\relax New York, NY, USA: Association for Computing Machinery, 2015, p.
  7¨C12.

\bibitem{UAV_LoS}
S.~{Zhang}, Y.~{Zeng}, and R.~{Zhang}, ``Cellular-enabled {UAV} communication:
  A connectivity-constrained trajectory optimization perspective,'' \emph{IEEE
  Transactions on Communications}, vol.~67, no.~3, pp. 2580--2604, 2019.

\bibitem{UAV_IoT}
Z.~{Wang}, R.~{Liu}, Q.~{Liu}, J.~S. {Thompson}, and M.~{Kadoch},
  ``Energy-efficient data collection and device positioning in {UAV}-assisted
  {IoT},'' \emph{IEEE Internet of Things Journal}, vol.~7, no.~2, pp.
  1122--1139, 2020.

\bibitem{Old_UAV_1}
C.~{Ou} and K.~{Ssu}, ``Sensor position determination with flying anchors in
  three-dimensional wireless sensor networks,'' \emph{IEEE Transactions on
  Mobile Computing}, vol.~7, no.~9, pp. 1084--1097, 2008.

\bibitem{Old_UAV_2}
G.~{Han}, J.~{Jiang}, C.~{Zhang}, T.~Q. {Duong}, M.~{Guizani}, and G.~K.
  {Karagiannidis}, ``A survey on mobile anchor node assisted localization in
  wireless sensor networks,'' \emph{IEEE Communications Surveys $\&$
  Tutorials}, vol.~18, no.~3, pp. 2220--2243, 2016.

\bibitem{UAV_HAWK}
Z.~{Liu}, Y.~{Chen}, B.~{Liu}, C.~{Cao}, and X.~{Fu}, ``{HAWK}: An unmanned
  mini-helicopter-based aerial wireless kit for localization,'' \emph{IEEE
  Transactions on Mobile Computing}, vol.~13, no.~2, pp. 287--298, 2014.

\bibitem{UAV_GuideLoc}
A.~Wang, X.~Ji, D.~Wu, X.~Bai, N.~Ding, J.~Pang, S.~Chen, X.~Chen, and D.~Fang,
  ``{GuideLoc}: {UAV}-assisted multitarget localization system for disaster
  rescue,'' \emph{Mobile Information Systems}, vol. 2017, 2017.

\bibitem{UAV_RSS_1}
H.~{Sallouha}, M.~M. {Azari}, A.~{Chiumento}, and S.~{Pollin}, ``Aerial anchors
  positioning for reliable {RSS}-based outdoor localization in urban
  environments,'' \emph{IEEE Wireless Communications Letters}, vol.~7, no.~3,
  pp. 376--379, 2018.

\bibitem{UAV_RSS_2}
H.~{Sallouha}, M.~M. {Azari}, and S.~{Pollin}, ``Energy-constrained {UAV}
  trajectory design for ground node localization,'' in \emph{2018 IEEE Global
  Communications Conference (GLOBECOM)}, 2018, pp. 1--7.

\bibitem{RAIM_Intr_1}
R.~G. Brown, ``Receiver autonomous integrity monitoring,'' \emph{Global
  positioning system: theory and applications}, vol.~2, pp. 143--165, 2003.

\bibitem{RAIM_SS}
R.~G. BROWN and P.~W. McBURNEY, ``Self-contained {GPS} integrity check using
  maximum solution separation,'' \emph{NAVIGATION}, vol.~35, no.~1, pp. 41--53,
  1988.

\bibitem{RAIM_PS}
M.~A. STURZA, ``Navigation system integrity monitoring using redundant
  measurements,'' \emph{NAVIGATION}, vol.~35, no.~4, pp. 483--501, 1988.

\bibitem{Multi_Fault_1}
M.~Joerger, F.-C. Chan, and B.~Pervan, ``Solution separation versus
  residual-based {RAIM},'' \emph{NAVIGATION}, vol.~61, no.~4, pp. 273--291,
  2014.

\bibitem{Multi_Fault_2}
J.~Blanch, A.~Ene, T.~Walter, and P.~Enge, ``An optimized multiple hypothesis
  {RAIM} algorithm for vertical guidance,'' in \emph{Proceedings of the 20th
  International Technical Meeting of the Satellite Division of The Institute of
  Navigation (ION GNSS 2007)}, 2007, pp. 2924--2933.

\bibitem{RAIM_NLoS}
N.~{Zhu}, J.~{Marais}, D.~{B¨¦taille}, and M.~{Berbineau}, ``{GNSS} position
  integrity in urban environments: A review of literature,'' \emph{IEEE
  Transactions on Intelligent Transportation Systems}, vol.~19, no.~9, pp.
  2762--2778, 2018.

\bibitem{LTE_MEDLL}
P.~{Wang} and Y.~J. {Morton}, ``Multipath estimating delay lock loop for {LTE}
  signal {TOA} estimation in indoor and urban environments,'' \emph{IEEE
  Transactions on Wireless Communications}, vol.~19, no.~8, pp. 5518--5530,
  2020.

\bibitem{TR_38.901}
{3GPP TR 38.901}, ``Study on channel model for frequencies from 0.5 to 100
  {GHz},'' Rel. 16, 2019.

\bibitem{Angle_based}
A.~{Al-Hourani}, S.~{Kandeepan}, and S.~{Lardner}, ``Optimal {LAP} altitude for
  maximum coverage,'' \emph{IEEE Wireless Communications Letters}, vol.~3,
  no.~6, pp. 569--572, 2014.

\bibitem{DEM_Rec}
D.~Kidner, M.~Dorey, and D.~Smith, ``What's the point? {Interpolation} and
  extrapolation with a regular grid {DEM},'' in \emph{Fourth International
  Conference on GeoComputation, Fredericksburg, VA, USA}, 1999.

\bibitem{TWR_Intr_2}
S.~Frattasi and F.~Della~Rosa, \emph{Mobile positioning and tracking: from
  conventional to cooperative techniques}.\hskip 1em plus 0.5em minus
  0.4em\relax John Wiley $\&$ Sons, 2017.

\bibitem{TWR_Intr_1}
C.~L. {Sang}, M.~{Adams}, T.~{H?rmann}, M.~{Hesse}, M.~{Porrmann}, and
  U.~{R¨¹ckert}, ``An analytical study of time of flight error estimation in
  two-way ranging methods,'' in \emph{2018 International Conference on Indoor
  Positioning and Indoor Navigation (IPIN)}, 2018, pp. 1--8.

\bibitem{TWR_Intr_3}
``{Part} 15.4: Wireless medium access control ({MAC}) and physical layer
  ({PHY}) specifications for low-rate wireless personal area networks
  ({WPANs}): {Amendment 1}: {Add} alternate {PHYs},'' \emph{IEEE Std
  802.15.4a-2007 (Amendment to IEEE Std 802.15.4-2006)}, pp. 1--210, 2007.

\bibitem{Max_Slope}
M.~Joerger and B.~Pervan, ``Kalman filter-based integrity monitoring against
  sensor faults,'' \emph{Journal of Guidance, Control, and Dynamics}, vol.~36,
  no.~2, pp. 349--361, 2013.

\end{thebibliography}

\end{document}